\documentclass[journal]{IEEEtran}

\usepackage[numbers,sort&compress]{natbib}
\usepackage{url}
\usepackage{bm}
\usepackage{amsfonts}
\usepackage{amsmath}
\usepackage{amssymb}
\usepackage{amsthm}
\usepackage{color}
\usepackage{enumerate}
\usepackage{makecell}
\usepackage{mathrsfs}
\usepackage{array,color}
\usepackage{tabularx}
\usepackage{multirow}
\usepackage{booktabs}
\usepackage{makecell}  

\usepackage{graphicx}  
\usepackage{subfig}
\usepackage{epstopdf}
\usepackage{graphics}
\usepackage{epsfig}
\usepackage{psfrag}
\usepackage{overpic}

\usepackage{multicol}

 \usepackage{stfloats}
 \usepackage{float}   

 \usepackage{indentfirst} 
\usepackage{gensymb}

\usepackage{color, colortbl}

\begin{document}

\title{Exploration of Superposition Theorem in Spectrum Space for Composite Event Analysis in an ADN}
\author{Xing~He,~\IEEEmembership{Senior Member},  Qian~Ai, Yuezhong~Tang,   Robert~Qiu,~\IEEEmembership{Fellow}, Canbing~Li
\thanks{This work was supported by Project Supported by National Science Foundation of China(52277111), and Science and Technology Commission of Shanghai Municipality(21DZ1208300).}
}
\maketitle

\begin{abstract}
This study presents a formulation of the Superposition Theorem (ST) in the spectrum space, tailored for the analysis of composite events in an active distribution network (ADN).
Our formulated ST enables a quantitative analysis on a composite event, uncovering the property of additivity among independent atom events in the spectrum space. This contribution is a significant addition to the existing literature and has profound implications in various application scenarios.
To accomplish this, we leverage random matrix theory (RMT), specifically the asymptotic empirical spectral distribution, Stieltjes transform, and R transform. These mathematical tools establish a nonlinear, model-free, and unsupervised addition operation in the spectrum space.
Comprehensive details, including a related roadmap, theorems, deductions, and proofs, are provided in this work.
Case studies, utilizing field data, validate our newly derived ST formulation by demonstrating a remarkable performance.
Our ST formulation is model-free, non-linear, non-supervised, theory-guided, and uncertainty-insensitive, making it a valuable asset in the realm of composite event analysis in ADN.
\end{abstract}


\IEEEpeerreviewmaketitle
\section{Introduction}
\label{Sec:Introd}
\IEEEPARstart{T}he term ``composite event'' refers to an event that is constructed from multiple atom events\footnote{Atom events are fundamental constituents of observable composite event. They are often directly invisible, relatively simpler, and mutually independent.}~\cite{gehani1992composite}.
The two-wave interference displayed in Figure~\ref{fig:DTVSModel} is a typical composite event.
The composite event is also pretty common in an active distribution network (ADN).
For instance, a composite event may be easily triggered by simultaneous power generation of distributed energy resources (DERs), which often exhibit diverse behaviors and considerable uncertainties~\cite{martins2011active}.

\subsection{Motivation of Our Works}

The presence of composite event may distort the typical pattern we observe, and hence disable tradition indicators we rely on, resulting in a poor (event-trigger) decision we make eventually.
This issue is particularly critical in today's ADN, which incorporates a multitude of DERs.

The composite event is analyzable, although it defies classical formulation, as discussed in Section~\ref{Sec: STEng}.
Triggered by its heterogeneous components (atom events), the composite event does induce some intricate yet identifiable (although maybe previously unknown) pattern, just akin to the two-wave interference pattern displayed in the right-bottom part of Figure~\ref{fig:DTVSModel}.
Such an induced pattern is usually accompanied by a nonlinear, stochastic, complex, and dynamic procedures, making it \textbf{too intricate for analytical solutions} or simple characterization through an a prior signature dictionary~\cite{wang2014MultipleEvent}.

Fortunately, numerous cutting-edge data technologies, such as advanced sensors, 5G/6G, and cloud platforms, have converged to supply massive structured (spatial-temporal) data from heterogeneous sensors, making a \textbf{numerical description as a viable alternative}.
It enable us to model the induced intricate pattern, in a straightforward manner, using the {so-called spatial-temporal data matrix}---a structured entirety with $N$ sampling points (spatial dimensions) and $T$ sampling times (temporal dimensions) each.
As a result, \textbf{we naturally turn the original composite event analysis task into jointly spatial-temporal analysis problem}.
High-dimensional statistics, more specifically, \textbf{random matrix theory (RMT) is proved to be effective to address this problem}~\cite{he2015arch}.

\subsection{Relevant Works}

Composite event analysis has demonstrated {fruitful} performance in many fields.
Reference~\cite{li2009delay}, in wireless sensor networks, devises an event detection model and a warning delivery model, with the goal of monitoring composite events and delivering warnings to users.
Reference~\cite{wang2010cooperative}, in image processing, proposes a wireless embedded smart-camera system for cooperative object tracking and detection of composite events spanning multiple camera views.
Reference~\cite{pitsikalis2019composite} develops a maritime monitoring system based on a Run-Time Event Calculus, and a composite event recognition system with formal, declarative semantics. 

\begin{figure*}[htbp]
\centerline{
\includegraphics[width=.88\textwidth]{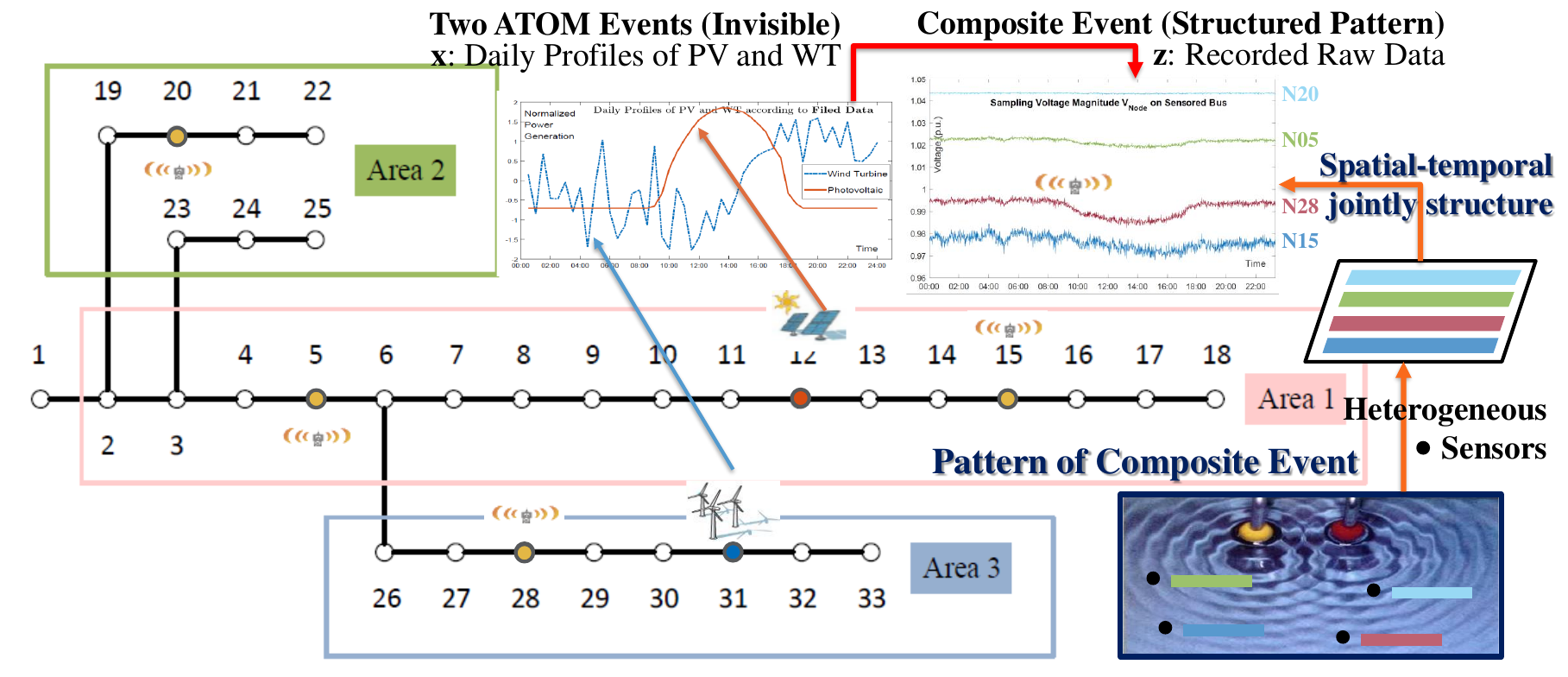}
}
\caption{Independently spatial-temporal analysis paradigm V.S. jointly spatial-temporal analysis paradigm}
\label{fig:DTVSModel}
\end{figure*}

Prevailing event analyses in power system, however, are mostly spatial-temporal disjointed, as illustrated at the left part of Figure~\ref{fig:DTVSModel}.
They are mainly based on physical/mechanical models with assumptions/simplifications, or a large labelled sample set, and thereby capable of handling ideal, typical, extreme, or frequent scenarios only.
Reference~\cite{Brahma2017Real} proposes a data-driven event identification method that can precisely determine the types of overlapping events, which is built upon a labelled phasor measurement unit (PMU) dataset for offline neural network training.
Reference~\cite{Tate2008Line} develops an algorithm that utilizes both PMU data and known system topology information to detect system line outages.
Reference~\cite{Biswa2016Supervisory, Yadav2019Real} study data-driven algorithms for multiple event detection and classification with a two-step process, in which the temporal analysis and spatial analysis are mutually independent.

Recently, the spatial-temporal jointed correlation attracted scholars' attention in power system, and big data analytics (BDA) became a stress and hot topic~\cite{bda2016tsg}.
Utilizing the sparsity property of signals, Reference~\cite{song2017MultipleEvent} presents a cluster-based sparse coding (CSC) for multiple event recognition, and Reference~\cite{wang2014MultipleEvent} presents a nonnegative sparse event unmixing (NSEU) algorithm, respectively.
Reference~\cite{Rafferty2019Real} study a data-driven method based on moving window principal component analysis (MW-PCA), allowing the spatial-temporal correlations to be taken into consideration.
Besides, to its inversion, decomposition of composite event, some NILM (non-intrusive load monitoring) solutions are discussed~\cite{kelly2015neural,  ciancetta2020new}.

Our previous work, concerning with spatial-temporal jointed correlation, has already made a series of exploration based on RMT.
{RMT is an emerging discipline closely associated with and taking a dominant role in high-dimensional statistics~\cite{qiu2015smart}}.
In Reference~\cite{he2015arch}, we firstly propose a unified and generalized RMT framework, which handles the spatial-temporal data as a matrix-variable (as depicted in the top part and the right part of Figure~\ref{fig:DTVSModel}).
The superiority and application scenarios of the proposed RMT framework are discussed in details in References~\cite{he2015corr, he2020hybrid}. This work is also an extension of our proposed RMT framework. See Section~\ref{Sec: STFor} for details.

\subsection{Contribution of Our Works}
\label{Sec:motiva}
This work strives to implement composite event analysis, a challenging task in an ADN.
In this regard, superposition theorem (ST) is formulated in spectrum space, providing \textbf{a quantitative analysis on and a numerical solution to} composite event analysis. Specially, this work, rooted in RMT, proposes a combination of \textbf{asymptotic empirical spectrum distribution and two transforms (Stieltjes and R transforms)}. The corresponding roadmap, theorems, deductions, and proofs are outlined to offer a comprehensive understanding of our work.

Our RMT-based composite event analysis is tailored to handle spatial-temporal data in the form of matrix-variables. Hence \textbf{it inherits several advantages associated with RMT}---it is model-free, non-supervised, theory-guided, and uncertainty-insensitive.  Consequently, it \textbf{outperforms in terms of complexity and noise accumulation}, making it highly competitive in engineering applications. Moreover, our method does not demand a large number of labeled data samples, a requirement that may often be impractical in real-world scenarios. Additionally, our approach is characterized by transparency due to its strong theoretical foundation.

Employing Stieltjes transform and R transform, this work introduces an abstract notation $\oplus$. This innovation, to the best of our knowledge, presents a pioneering exploration into the \textbf{additivity properties of atom events within the spectrum space}. This groundbreaking finding paves the way for advanced applications in various domains, including but not limited to component identification and anomaly detection.
This work could be helpful not only to ADN dispatchers, who may encounter obsolete topology configuration information due to reconfiguration of ADN network (with wrong physical model information), but also to VPP operators, who may only have local knowledge confined to their region (with no real-time model information).

Our work is organized as follows:
Section~\ref{Sec: STEng} revisits traditional Superposition Theorem (ST) formulations for event superposition.
Section~\ref{Sec: STFor}, by employing {jointly spatial-temporal analysis}, reformulates ST in spectrum space.
Section~\ref{Sec: Case} studies some common scenes in ADN to verify our suggested composite event analysis.
Section~\ref{Sec: Conc} concludes our work.

\section{{Composite Event Formulations in Power Grid}}
\label{Sec: STEng}

{It is well-known that ``linear addition +'', ``model-based addition $\boxplus$'', and ``Fourier addition $\uplus$''
are used to formulate concurrent events superposition, respectively, in circuit analysis, power flow analysis (PFA),  and harmonic analysis.
Based on these addition rules, many advanced servers/functions are further developed, e.g., line loss calculation~\cite{sun1980calculation}, harmonic source detection~\cite{sinvula2019harmonic}.
In contrast to the above addition, this work strives to define an addition ``$\oplus$'' \textbf{concerning with the formulation of  heterogeneous DERs' influence upon ADN}, as shown in Figure~\ref{fig:DTVSModel}.
This exploration should be very helpful for ADN dispatchers and VPP operators.}

\subsection{Linear Addition $+$ with Traditional Superposition Theorem}
\begin{figure}[ht]
\centering
\subfloat[{Classical ST: Linear addition $+$ }]{\label{fig:ClassicalST}
\includegraphics[width=0.46\textwidth]{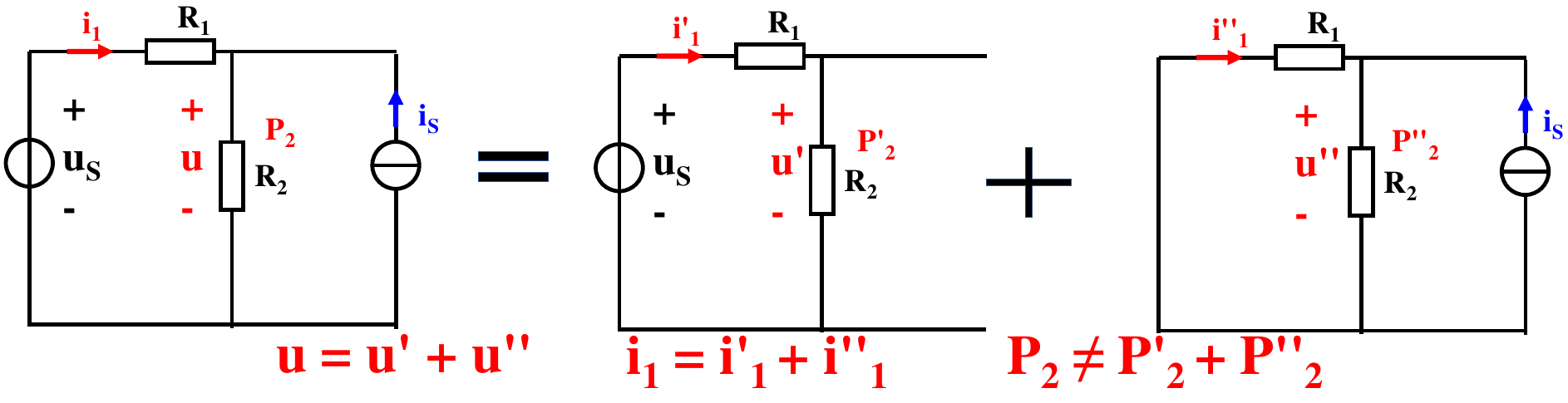}}

\subfloat[Th\'evenin Conversion]{\label{fig:Theve}
\includegraphics[width=0.26\textwidth]{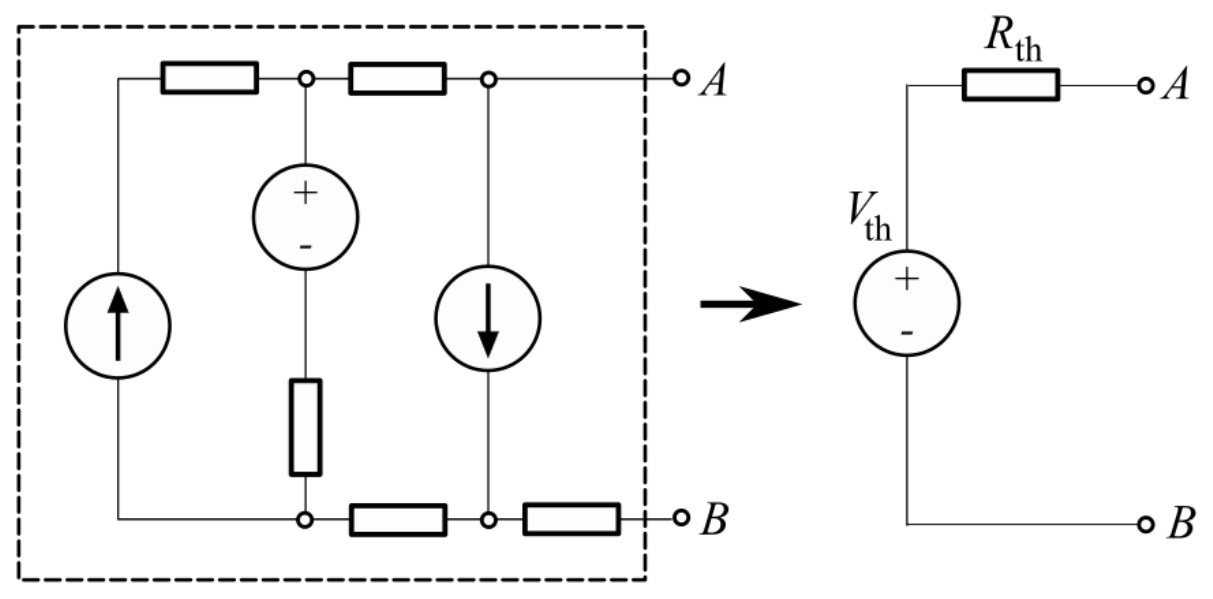}}
\subfloat[Norton-Th\'evenin Conversion]{\label{fig:Norton}
\includegraphics[width=0.23\textwidth]{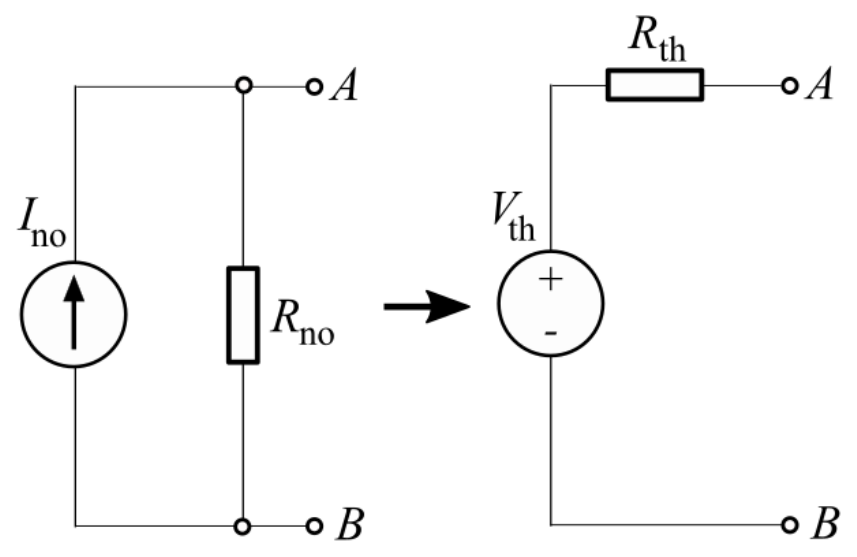}}
\caption{Any black box containing only resistances, and voltage and current sources can be replaced by a Th\'evenin equivalent or Norton equivalent circuit consisting of an equivalent voltage source in series connection with an equivalent resistance.}
\label{fig:ConvenST}
\end{figure}

Superposition Theorem (ST), as shown in Figure~\ref{fig:ClassicalST}, is one of the most fundamental theorem in conventional circuit analysis. It tells that for a linear system, the response (voltage or current) in any branch of a bilateral linear circuit having more than one independent source equals the algebraic sum of the responses caused by each independent source acting alone, where all the other independent sources are replaced by their internal impedances~\cite{svoboda2013introduction}.

ST helps to solve linear circuit with more than one current or voltage source, outputting the resultant linear parameters including voltage $U$ and current $I$.
It could aid in converting any linear circuit into its Th\'evenin equivalent or Norton equivalent, as shown in Figure~\ref{fig:ConvenST}.
However, ST is only applicable for linear parameters but not for non-linear ones such as power $P, Q$, let alone the unanticipated trigged event.
{In summary, traditional ST defines a {``linear addition $+$''}  for linear parameters $U, I$.}

\subsection{Model-based Addition $\boxplus$ in Power Flow Analysis}
\label{Sec: STEng-B}
For non-linear factors  $P, Q$, we switch to PFA.
PFA states one of the most fundamental equations for describing the grid's behavior in the steady-state \cite{gomez2018electric}.
PFA says that for each node, Node $i$ for instance,
\begin{equation}
\small
\label{eq:PQend}
\begin{aligned}
  & \left\{ \begin{aligned}
  & {{P}_{i}}\!=\!{{V}_{i}}\sum\limits_{k\ne i}{{{V}_{k}}\left( {{\Psi}_{ik}}\text{cos}{{\theta }_{ik}}\!+\!{{\Phi}_{ik}}\text{sin}{{\theta }_{ik}} \right)}\!-\!{{V}_{i}}^{2}\sum\limits_{k\ne i}{{{\Psi}_{ik}}} \\
 & {{Q}_{i}}\!=\!{{V}_{i}}\sum\limits_{k\ne i}{{{V}_{k}}\left( {{\Psi}_{ik}}\sin {{\theta }_{ik}}\!-\!{{\Phi}_{ik}}\text{cos}{{\theta }_{ik}} \right)}\!+\!{{V}_{i}}^{2}\sum\limits_{k\ne i}{{{\Phi}_{ik}}} \\
\end{aligned} \right. \\
 &      \\
\end{aligned}
\normalsize
\end{equation}

Abstractly, Eq.~\eqref{eq:PQend} can be regarded as a model-based analog engine---it takes bus voltage magnitude $V$ and phase {angle} $\theta$ as {inputs}, conductance $\Psi$ and susceptance $\Phi$ as given parameters, and ``computes''  active power injection $P$ and reactive power injection $Q$  as {outputs}.
Many classical iteration methods, e.g.,  Newton-Raphson algorithms~\cite{da1999developments}, are capable of solving Eq.~\eqref{eq:PQend}.
It is worth noting that the prior knowledge about each physical parameter (e.g., $\Psi, \Phi$ in Eq.~\ref{eq:PQend}) is required for this formulation.
{In summary, PFA defines a {``model-based addition $\boxplus$''}  for  non-linear parameters $P, Q$.}

\subsection{{Fourier Addition $\uplus$ in Harmonic Analysis}}

{Harmonic analysis is concerned with the representation of observed signals as the superposition of some particular waves in a power system.
Harmonic analysis is built upon some basic assumptions~\cite{85861}: a) The signal is stationary (constant magnitude); b) The sampling frequency is greater than twice the highest frequency in the signal to be analyzed; and c) Each frequency in the signal is an integer multiple of the fundamental frequency $\omega$. (The presence of interharmonic components may largely decrease the performance~\cite{4302786}.)
When these assumptions are satisfied, we have}
\begin{equation}
\label{eq:Harmonic}
s(t)=\sum_{i=1}^n A_i(t) \cos \left(i \omega t+\theta_i\right)
\end{equation}
{where
$A_i(t)$ is the amplitude of the phasor quantity representing the $i^{\text {th }}$ harmonic at time $t$,
$\theta_i$ is the phase angle of the $i^{\text {th }}$ harmonic relative to a reference rotating at ${i} \omega$,
and ${n}$ is the harmonic order.
In summary, Fourier transform  defines a ``Fourier addition $\uplus$'' in frequency-series space.
}

\subsection{Our Addition Formulation $\oplus$ in High-dimensional Space}
\label{sec: MovitationRMT}

{The composite event in ADN, as illustrated in Figure~\ref{fig:DTVSModel}, is often non-linear and without reliable topology knowledge, especially for those grid dispatchers or VPP operators.
The addition of such concurrent events cannot be effectively formulated using above ``linear addition +'', ``model-based addition $\boxplus$'', or ``Fourier addition $\uplus$''.
Fortunately, the ever-increasing sampled spatial-temporal data, which already cover the majority of the composite event, can help inspect and conceptualize the composite event and its intricate yet identifiable pattern.
}
{
Those spatial-temporal data naturally form data matrix, a structured entirety with $N$ sampling points (spatial dimension) and $T$ sampling times (temporal dimension) each.
As a result, high-dimensional statistics, more specifically, random matrix theory (RMT), is tightly tied to our formulation. Furthermore, we strive to formulate some ``RMT-derived addition $\oplus$'', as}
\begin{equation}
\label{eq:STAbs}
f(\mathbf M)=f(\mathbf A_1)\oplus f(\mathbf A_2)\oplus\cdots,
\end{equation}
where $\mathbf M, \mathbf A_i \!\in\! {{\mathbb{R}}^{N\times T}}$. $\mathbf M$ is the random matrix for the resultant composite event, and $\mathbf A_i$ for each of its components, i.e.,  independent atom event each. {$f(\cdot)$ is some RMT-derived function that projects the (event data) matrix onto a (high-dimensional) variable}.
In practice,  $\mathbf M$ is usually observable directly, while $\mathbf A_i$ is invisible.

In this way, Eq.~\eqref{eq:STAbs} tries to {formulate event superposition phenomenon as a matrix-variables operation}---the superposition of concurrent atom events (e.g., two Atom Events in Figure~\ref{fig:DTVSModel}) is turned into  ``RMT-derived addition $\oplus$'' of their corresponding matrix-variables.
\textbf{The key challenge lies in defining the operation symbol ``$\oplus$''}.
{Throughout our study, this challenge remains a central theme, and we provide a roadmap in Figure~\ref{fig:GRsum}}.

\section{RMT-derived Addition in Spectrum Space}
\label{Sec: STFor}
As discussed in Section~\ref{Sec:Introd}, RMT tools  are naturally connected to composite event analysis task.


{\subsection{RMT in Engineering}}

{
In engineering, these raw data are sampled by heterogeneous sensors independently, and then gather together in the data platform/cloud.
Traditionally, they are used in a spatial-temporally disjointed manner---either in the form of an isolated time-series, or in the form of a single time slice.}

{
Jointly spatial-temporal analysis is a challenging task in modern statistics~\cite{7587390}---it is hard for spatial-temporal data to get correlated to a unified and generalized framework.
This is the motivation for our research on RMT, which is naturally connected to spatial-temporal data.
To our best knowledge, RMT is developed to address the high-dimensional regime since the classical statistic theory applies to the low-dimensional regime only~\cite{qiu2015smart}.
Under the RMT framework~\cite{he2015arch}, we often encounter a dataset with {a big number of observed variables} $N$ and {a large number of  sampling points} $T$, but a ratio $c \!=\! N/T$ that is not small compared to unity. This setting is known as the high-dimensional limit in the literature; it is totally different from the traditional large $T$, fixed $N$ (small, often $N\!<\!6$~\cite{qiu2015smart}) situation (i.e. $c\!\to\! 0$).
}

{RMT, concerning with the joint distribution of eigenvalues as the statistic analytics, mainly studies the eigenvalues (spectrum) of covariance matrices}, an object of central interest in multivariate (high-dimensional) statistics~\cite{elkaroui2008}.
{The spectrum has strong correlation with principal/independent/free component analysis  (PCA/ICA/FCA)~\cite{wu2019free}, where one tries to find out the ``best'' possible component from some observed intricate pattern}, and thus can be further used for anomaly detection~\cite{he2015arch} or event disaggregation~\cite{he2020hybrid}.

\subsection{Asymptotic Empirical Spectral Distribution based on RMT}

In RMT, Gaussian unitary ensemble (GUE) and Laguerre unitary ensemble (LUE) are studied firstly~\cite{he2018SA}:
\begin{equation}
\label{eq:B1}
\mathbf{\Gamma}=\left\{ \begin{aligned}
  & \frac{1}{2\sqrt{N}}\left( \mathbf{C}+{{\mathbf{C}}^{H}} \right)&, \mathbf{C}\in {{\mathbb{C}}^{N\times N}}& \text{  ,GUE;} \\
 & \frac{1}{T}\mathbf{R}{{\mathbf{R}}^{H}}&, \mathbf{R}\in {{\mathbb{R}}^{N\times T}}& \text{  ,LUE}\text{.} \\
\end{aligned} \right.,
\end{equation}
where $\mathbf{R}$ is a standard Gaussian random matrix whose entries are i.i.d. real Gaussian random variables, and $\mathbf{C}$ is a complex Gaussian random matrix whose entries are i.i.d. complex Gaussian random variables ($\mathbf{C} = \mathbf{R_1}+i\cdot\mathbf{R_2}$, where $\mathbf{R_1}, \mathbf{R_1}$, are two independent standard Gaussian random matrices.)

Let ${f_{\mathbf{\Gamma}}}\left( x \right)$ be the empirical density of $\mathbf{\Gamma}$, and define its empirical spectral distribution (ESD) ${F_{\mathbf{\Gamma}}}\left( x \right)$:
\begin{equation}
{F_{\mathbf{\Gamma}}}\left( x \right) = \frac{1}{N}\sum\limits_{i = 1}^N {{I_{\left\{ {{\lambda _i} \le x} \right\}}}},
\end{equation}
where $\mathbf{\Gamma}$ is GUE (or LUE) matrix, and $I\left(  \cdot  \right)$ represents the event indicator function. We investigate the rate of convergence of the expected ESD $\mathbb{E}\left( {{F_{\mathbf{\Gamma}}}\left( x \right)} \right)$ to the Wigner's Semicircle Law (or Wishart's M-P Law).

Let ${h_{\mathbf{\Gamma}}}\left( x \right)$ and ${H_{\mathbf{\Gamma}}}\left( x \right)$ denote the true eigenvalue density and the true spectral distribution of $\mathbf{\Gamma}$ in Eq.~\eqref{eq:B1}. For GUE and LUE, respectively, the Wigner's Semicircle Law and Wishart's M-P Law say that

\begin{equation}
\label{eq:D2}
{h_{\mathbf{\Gamma}}}\left( x \right) = \left\{ \begin{aligned}
&\frac{1}{{2\pi }}\sqrt {4 - {x^2}} &, x \in \left[ { - 2,2} \right]&\text{  ,GUE}; \\
&\frac{1}{{2\pi cx}}\sqrt {\left( {a_2 - x} \right)\left( {x - a_1} \right)} &, x \in \left[ { a_1,a_2} \right] &\text{  ,LUE};
\end{aligned} \right.,
\end{equation}
where $a_1 = {\left( {1 - \sqrt c  } \right)^2}, a_2 = {\left( {1 + \sqrt c  } \right)^2}$, and $c=N/T\le1$.

\[
{H_{\mathbf{\Gamma}}}\left( x \right) = \int_{ - \infty }^x {{h_{\mathbf{\Gamma}}}\left( u \right)\text{d}u} .
\]
Then, we denote the Kolmogorov distance between $\mathbb{E}\left( {{F_{\mathbf{\Gamma}}}\left( x \right)} \right)$ and ${H_{\mathbf{\Gamma}}}\left( x \right)$ as  $\Delta$:
\[
\Delta  = \mathop {\operatorname{sup}}\limits_x \left| {\mathbb{E}\left( {{F_{\mathbf{\Gamma}}}\left( x \right)} \right) - {H_{\mathbf{\Gamma}}}\left( x \right)} \right|.
\]
Gotze and Tikhomirov, in~\cite{gotze2005rate}, prove an optimal bound for $\Delta$  of order $O\left( {{N^{ - 1}}} \right)$.
It means that {more heterogeneous  sources in engineering may theoretically benefit our analysis}.
This is a very reason why RMT is suitable for spatial-temporal data.
In practice,  RMT performs well with (unlabelled) data in {moderate size  such as tens}, which is often true for ADN.

\subsection{Stieltjes Transform of Random Matrix}

Stieltjes Transform, built upon the spectral distribution of matrices~\cite{elkaroui2008}, enables the event addition, i.e., $f(\mathbf A_i)\oplus f(\mathbf A_j)$ in Eq.~\ref{eq:STAbs}.
{Stieltjes transform is closely connected with RMT}.
A large number of results concerning the asymptotic properties of the eigenvalues of large dimensional random matrices are formulated in terms of limiting behavior of the Stieltjes transform of their ESD.

\subsubsection{Random Matrix and its Spectrum Space}
{\text{\\}}

Consider a Hermitian random matrix $\bf X$.
For a continuous function $\varphi$, we can built a mapping by defining  normalized trace (i.e. $\textrm{tr}\left( {\mathbf{I}_N}\right)=1$,  $\mathbf{I}_N\in {{\mathbb{R}}^{N\times N}}$ represents the identity matrix) when $N \to \infty$, since that $\textrm{tr}\left( \varphi(\mathbf A)\right)\!=\!\sum_{t} \varphi(\lambda_{\mathbf A, t})$
\begin{equation}
\label{eq:tr_X}
\lim _{N \to \infty} \mathbb{E}\left({\textrm{tr}\left( \varphi(\mathbf X) \right)}\right) \!= \! \lim _{N \to \infty} \mathbb{E}\left(  \sum_{t=1}^{N} \varphi(\lambda_{\mathbf X, t}) \right)\! =\!  \int\limits_\mathbb{R} {\varphi(s)} \text{d}v(s),
\end{equation}
where $\lambda_{\mathbf X, t}$ is the $t$-th eigenvalue of $\bf X$.
And $v(s)$ denotes the spectrum distribution of $\mathbf X,$
\begin{equation}
\label{eq:tr_X2}
\text{d}v(s):=\rho(s)\text{d}s
\end{equation}
where $\rho(s)$ is the eigenvalue density at  $\lambda \!= \!s.$
Note that $\lambda_{\varphi(\mathbf X)}=\varphi(\lambda_{\mathbf X}).$

\subsubsection{Definition of Stieltjes Transform}
{\text{\\}}

Stieltjes transform is a convenient and powerful tool in the study of the convergence of spectral distribution of matrices. Just as the characteristic function of a probability distribution for central limit theorems, i.e., $\varphi_{X}(t)\!=\!e^ {{i} tX}$,
Stieltjes transform is specially useful to study the limit spectral properties and to {tackle the polynomial calculation of random matrices}.
Based on it,  R transform is suggested by problems of interest in engineering \cite{qiu2015smart}.

\textit{Definition} (Stieltjes Transform).
Let $v$ be a non-negative, finite Borel measure on $\mathbb{R}$ and for $z\!\notin\! \mathbb{R}$. Then Stieltjes transform of $v$ is defined as
\begin{eqnarray}
\label{eq:DefStj}
{G_v}(z) = \int\limits_\mathbb{R} {\frac{1}{{z - s}}} \text{d} v(s) = \int\limits_\mathbb{R} {\frac{\rho(s)}{{z - s}}} \text{d}s.
\end{eqnarray}
For all $z \in \{z:z \in \mathbb{C}, \textrm {Im} (z)>0\}$.

In  Eq.~\eqref{eq:tr_X}, we take $ \varphi(\mathbf x) = \left(z - x\right)^{-1}$.
As a result, it is deduced that Stieltjes transform is actually simplified to
\begin{equation}
\label{eq:Transformss}
 {G}(z) =\lim _{N \to \infty} \mathbb{E}\left({\textrm{tr}\left( z{{\mathbf{I}_N} - {\bf{X}}} \right)^{-1}}\right),
\end{equation}

\subsubsection{Stieltjes Transform of GUE and LUE Ensemble}
{\text{\\}}

Here, we give the analytical solution of Stieltjes transform of GUE and LUE by combining Eq.~\eqref{eq:D2} and~\eqref{eq:DefStj}.
\begin{equation}
\label{eq:STJtrans}
{G_{\mathbf{\Gamma}}}\left( z \right) = \left\{ \begin{aligned}
&\frac{z-\sqrt {4 - {z^2}}}{{2 }} &, \ \text{GUE} \\
&\frac{z+1-c-\sqrt {\left( {z - a_1} \right)\left( {z - a_2 } \right)}}{2z} &, \ \text{LUE}\\
\end{aligned} \right.,
\end{equation}
where $a_1 = {\left( {1 - \sqrt c  } \right)^2}, a_2 = {\left( {1 + \sqrt c  } \right)^2}$, and $c=N/T\le1$.
The proof is given in Appendix~\ref{Sec:ProofofStj}.

\subsubsection{Validation of Stieltjes Transform}
{\text{\\}}

\begin{figure}[h]
\centering
\subfloat[Imag part of ${G_v}(z)$]{\label{fig:stjreal}
\includegraphics[width=0.46\textwidth]{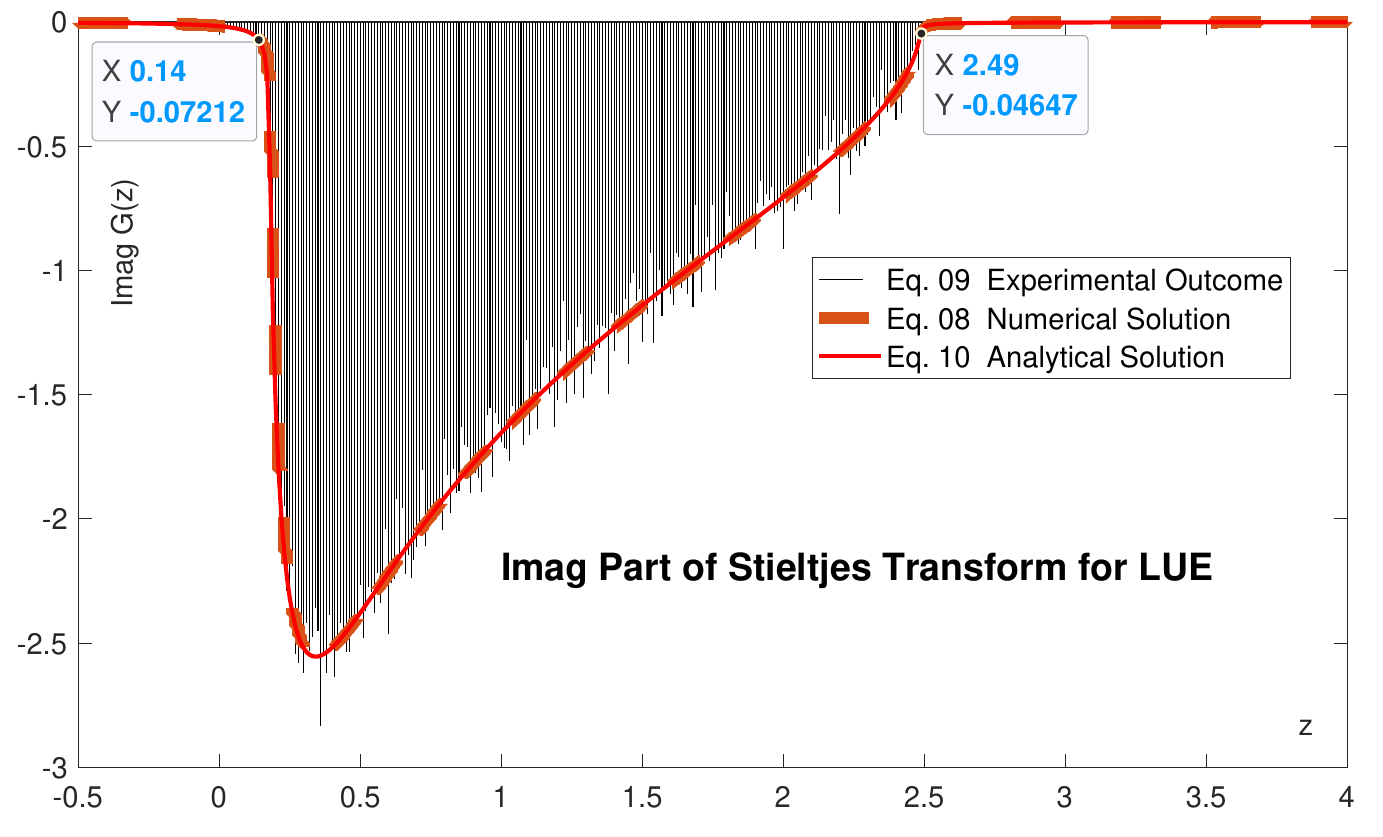}}

\subfloat[Real part of ${G_v}(z)$]{\label{fig:stjimag}
\includegraphics[width=0.46\textwidth]{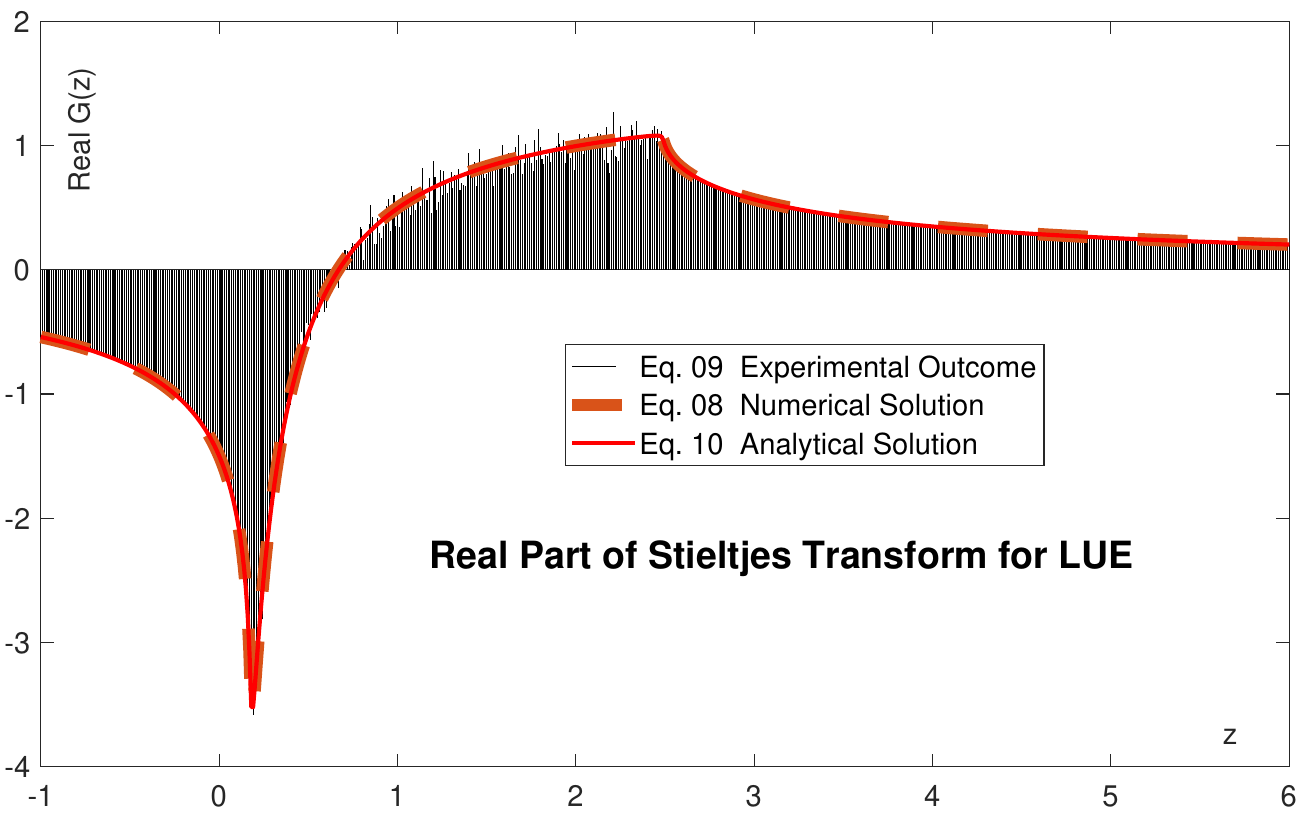}}
\caption{Stieltjes transform of LUE when $N=1000, T=3000 (c = 0.3333, a = 0.1786, b = 2.4880)$.  The experimental outcome (Eq.~\ref{eq:Transformss}), the numerical solution (Eq.~\ref{eq:DefStj}), and the analytical solution (Eq.~\ref{eq:STJtrans}) are perfectly matched.}
\label{fig:stj}
\end{figure}

Figure~\ref{fig:stj} depicts Stieltjes transform of LUE.
\begin{itemize}
  \item The numerical solution (thick dash line) is acquired by substituting  $\rho(s)$ in Eq.~\eqref{eq:DefStj} according to Eq.~\eqref{eq:D2}, i.e.,  $\rho(s)=\frac{1}{{2\pi cs}}\sqrt {\left( {s - a_1} \right)\left( {a_2 - s} \right)}, s \in \left[ { a_1,a_2} \right]$.
  \item The experimental outcome (stem thin line) is obtained through substituting  $\mathbf X$ in Eq.~\eqref{eq:Transformss} according to Eq.~\eqref{eq:B1}, i.e.,   $\mathbf X =\frac{1}{T}\mathbf{R}{{\mathbf{R}}^{H}}, \mathbf{R}\in {{\mathbb{R}}^{N\times T}}$.
  \item The analytical solution (thin solid line) is computed via Eq.~\eqref{eq:STJtrans} theoretically.
\end{itemize}

{The three lines are perfectly matched} on both image part and real part.
In this way, RMT and R Transform are naturally connected to our task composite event analysis in ADN.
We try to complete Eq.~\eqref{eq:STAbs}, by designing spectrum function $f(\cdot)$ and abstract addition $\oplus$,  in order to turn the spatial-temporal data into computable matrix-variables in the spectrum space.

Furthermore, we give the Stieltjes Inversion Formula as
\begin{equation}
\label{Transforms:InvG}
\rho(x) = -\frac{1}{\pi} \operatorname{Im}(G(x+i \epsilon)),  \   \epsilon\to 0^{+}
\end{equation}

\begin{figure}[htbp]
\centerline{
\includegraphics[width=.46\textwidth]{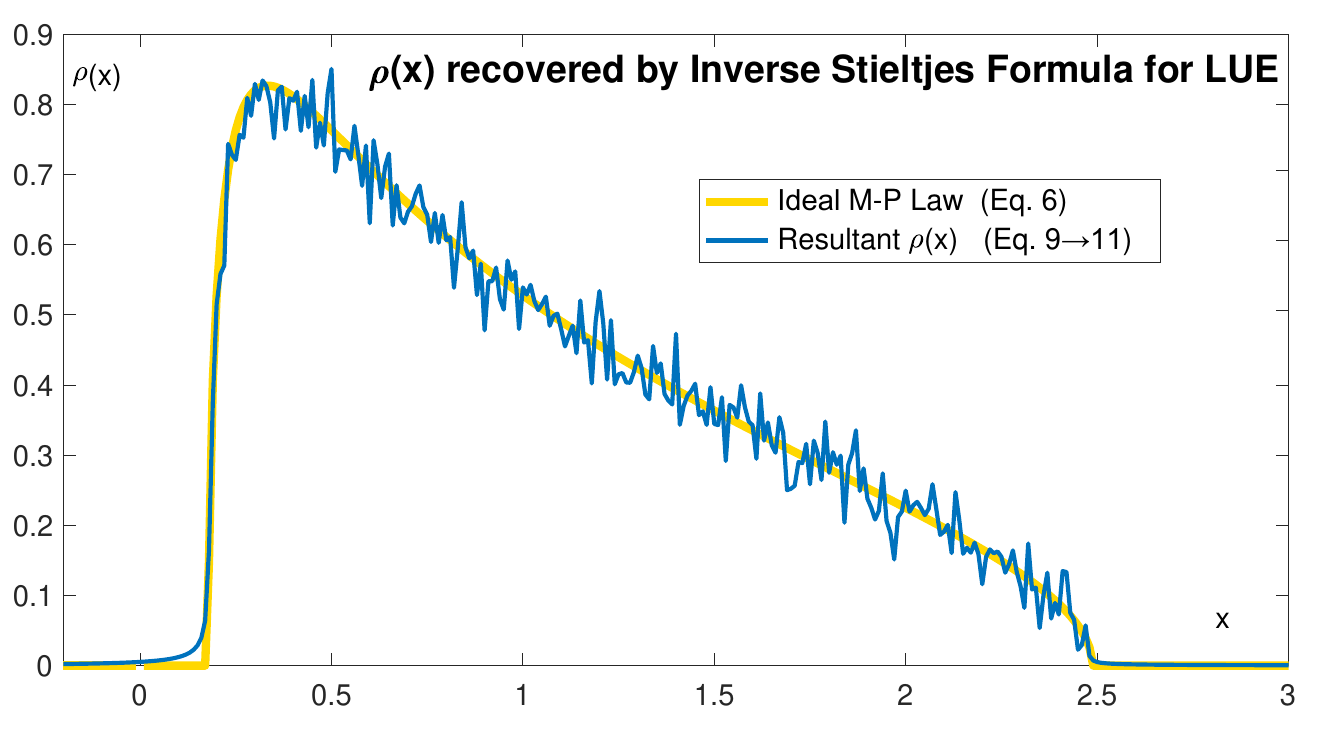}
}
\caption{Inverse Stieltjes transform of LUE}
\label{fig:invSjt}
\end{figure}

Inverse Stieltjes Formula is used to recover the  eigenvalue density $\rho(x)$ in the spectrum space.
Figure~\ref{fig:invSjt} shows that the experimental outcome $\rho(x)$ (Eq.~\ref{eq:Transformss} $\rightarrow$ \ref{Transforms:InvG})  is entirely consistent with M-P Law (Eq.~\ref{eq:D2}).

\subsection{Additivity Law (R) for Event Addition}

$R$ Transform is derived from Stieltjes transform.
$R$ Transform  enables the characterization of the limiting spectrum of a {sum} of random matrices from their individual limiting spectra.
These properties would turn out to be extremely useful in the engineering system whose data in the form of spatial-temporal rectangular matrix $\{\mathbf A_1, \mathbf A_2, \cdots\}, \mathbf A_i\! \in\! \mathbb R^{N\!\times\! T}$.
These matrix-variables $\mathbf A_i$ can be further turned into a Hermitian random matrix $\mathbf X_i$: $\mathbf X_i = 1/T\mathbf A_i \mathbf A_i^{\text T}$, where $\mathbf A_i^{\text T}$ is the transpose matrix of $\mathbf A_i$ and $\mathbf X_i\! \in\! \mathbb R^{N\!\times\! N}$.

For a Hermitian matrix $\mathbf X$, its $k$-th moment is defined as
\begin{equation}
\label{eq:moment_mk}
m_{k}=\mathbb{E}\left( \text{tr}\left( {{\mathbf{X}}^{k}} \right) \right)=\int{{{s}^{k}}\text{d}v\left( s \right)}
\end{equation}

Following we define the moment series of $\mathbf X$
\begin{equation}
\label{eq:momentSeries_Mz}
\xi\left( z \right):=\sum\limits_{n=0}^{\infty}{{{m}_{n}}{{z}^{n}}}=\sum\limits_{k=0}^{\infty}{\int\limits_{\mathbb{R}}{{{\left( zs \right)}^{k}}\text{d}v(s)}}
\end{equation}

and cumulant series
\[
\label{eq:cumulantSeries_Mz}
\zeta\left( z \right):=\sum\limits_{n=0}^{\infty}{{{\kappa}_{n}}{{z}^{n}}}
\]
where $
{{m}_{n}}\!=\!\sum\limits_{\pi \in \mathcal{P}\left( \left[ n \right] \right)}{{{\kappa }_{\pi }}}\!=\!{{B}_{n}}\left( {{\kappa }_{1}},\ldots ,{{\kappa }_{n}} \right)
.$
Note that $\mathcal{P}\left( \left[ n \right] \right)$ denote all \textit{partitions} of the set $[n]$, and ${{B}_{n}}\left( {{\kappa }_{1}},\ldots ,{{\kappa }_{n}} \right)$ is the Bell polynomials. (see Appendix~\ref{Sec:Bell}).

\textit{Proposition}. The relation between the moment series $M(z)$ and the cumulant series $K(z)$ is given by~\cite{Mingo2017Free}
\begin{equation}
\label{eq:MomentSeriesCumulantSeries}
\xi(z)=1+\zeta(z \xi(z))
\end{equation}
see Appendix~\ref{Sec:MzKz} for the proof.

Now consider the Stieltjes Transform $G$:

\begin{equation}
\label{eq:GM}
\begin{aligned}
   {{G}}(z)&=\int\limits_{\mathbb{R}}{\frac{1}{z-s}}\text{d}v(s)=\int\limits_{\mathbb{R}}{\frac{1}{z}\frac{1}{1-s/z}}\text{d}v(s) \\
 & =\int\limits_{\mathbb{R}}{\frac{1}{z}\sum\limits_{k=0}^{\infty}{{{\left( \frac{s}{z} \right)}^{k}}}}\text{d}v(s)=\frac{1}{z}\sum\limits_{k=0}^{\infty}{\int\limits_{\mathbb{R}}{{{\left( \frac{s}{z} \right)}^{k}}\text{d}v(s)}} \\
 & =\sum\limits_{k=0}^{\infty} \frac{m_n}{z^{k+1}} =\frac{1}{z}\xi\left( \frac{1}{z} \right)
\end{aligned}
\end{equation}

The $R$ Transform is defined by
\begin{equation}
\label{eq:eqRG}
R(z):=\frac{\zeta(z)}{z}
\end{equation}

And furthermore, if we define
\begin{equation}
\label{eq:T_invG}
T(z):=G^{ - 1}(z)
\end{equation}

it is obtained that
\begin{equation}
\label{eq:RfromT}
{R}(z) =T(z)-\frac{1}{z}
\end{equation}
See Appendix~\ref{Sec:ProofofG-1} for the proof.

\subsubsection{Properties of $R$ Transform}
{\text{\\}}

Then we give two important properties of $R$ Transform:

1.  $\bf{Additivity \ Law}$:
let ${R_{\bf{A}}}\left( z \right)$, ${R_{\bf{B}}}\left( z \right)$ and ${R_{\bf{A}+\bf{B}}}\left( z \right)$ be the $R$ Transform of matrices $\bf{A}$, $\bf{B}$ and $\bf{A}+\bf{B}$, respectively. We have
\begin{equation}
\label{Transforms:eq6}
{R_{{\bf{A} + \bf{B}}}}\left( z \right) = {R_{\bf{A}}}\left( z \right) + {R_{\bf{B}}}\left( z \right).
\end{equation}

2.  $\bf{Scaling \ Property}$:
For any $\alpha>0$,
\begin{equation}
\label{Transforms:eq7}
{R_{\alpha {\bf{X}}}}\left( z \right) = \alpha {R_{\bf{X}}}\left( {\alpha z} \right).
\end{equation}

Additivity law can be easily understood in terms of Feynman diagrams (see~\cite{qiu2015smart,voiculescu1992free} for details). The above two properties of $R$ Transform enable us to conduct {linear calculation} of the asymptotic spectrum of random matrices.

\begin{figure}[!ht]
  \centering
  \includegraphics[width=0.46\textwidth]{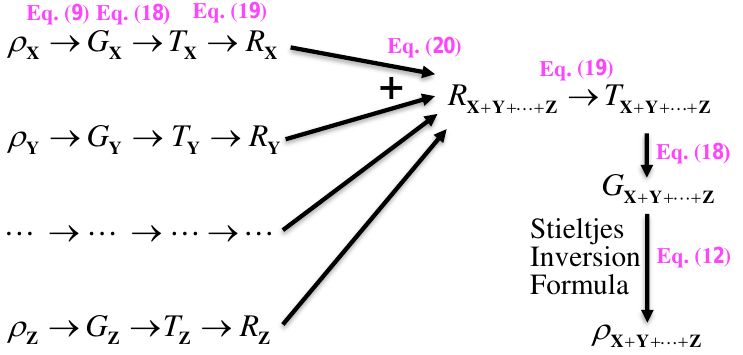}
  \caption{Graphical illustration of our multi-events operation.  The $\oplus$ is defined in spectrum space.}
  \label{fig:GRsum}
\end{figure}

With Stieltjes Transform ($G$ Transform)  and $R$ Transform, asymptotic free independence of {the sum} of elements $\mathbf X_i\mathbf X_i^\text{H}$ is acquired. In such a way, we define an {abstract addition $\oplus$ to complete Eq.~\eqref{eq:STAbs} in the spectrum space}.
The general idea of implementing multi-events operation is given in Figure~\ref{fig:GRsum}.

In such a manner, we model the phenomena of a composite event in spectrum space by defining an abstract addition $\oplus$.
Unlike raw data or most classical low-dimensional statistics, the spectrum does depend on the entirety of a matrix.
Hence spectrum-derived indicators embody much more information on spatial-temporal correlation, and their operation are usually sensitive to signals and robust against i.i.d. noise~\cite{he2015arch}.

\subsubsection{Example of Additivity Law (R)}
{\text{\\}}

Let \[{{\mathbf Q}_n}\!=\!\frac{1}{n} {{{\mathbf X}_{1}}\mathbf X_{1}^{'}}\!+\!\cdots \!+\!\frac{1}{n} {{{\mathbf X}_{k}}\mathbf X_{k}^{'}}\]
where $\mathbf X_i \! \sim\! {{\mathscr N}_{p,p}}(0,{\mathbf I_p},{\mathbf I_p})$ for $i=1,\cdots,k$. $\mathbf  {X}_i$ and $\mathbf  {X}_j$ are independent for all $i\neq j$. Then, the asymptotic eigenvalue distribution of $\mathbf  {Y}_i=\frac{1}{p}{\mathbf {X}_{i}} \mathbf {X}_{i}^{'}$ follows M-P Law:
\[{\rho}(x)=\frac{1}{{2\pi x}}{\sqrt {4x - {x^2}} }, \quad  x \in \left[ { 0,4} \right]  \qquad \Leftarrow  \text{Eq.~\eqref{eq:D2}}, c=1 \]
Then,
\[{G_{{\bf Y}_i}}(z) = \frac{1}{2} - \sqrt {\frac{1}{4} - \frac{1}{z}}, \qquad \Leftarrow \text{Eq.~\eqref{eq:STJtrans}} \]
\[T_{{\bf Y}_i}(y) = G_{{\bf Y}_i}^{ - 1}(y) = \frac{1}{{(1 - y)y}}, \qquad \Leftarrow  \text{Eq.~\eqref{eq:T_invG}} \]
\[{R_{{\bf Y}_i}}(z) = \frac{1}{{(1 - z)z}} - \frac{1}{z} = \frac{1}{{(1 - z)}}, \qquad \Leftarrow  \text{Eq.~\eqref{eq:RfromT}}\]
Hence according to Additivity Law
\[{R_{{{\bf Q}_n}}}(z) = \frac{k}{{1 - z}}, \qquad \Leftarrow  \text{Eq.~\eqref{Transforms:eq6}}\]
 \[T_{{{\bf Q}_n}}(z) = \frac{k}{{1 - z}} + \frac{1}{z}, \qquad \Leftarrow  \text{Eq.~\eqref{eq:T_invG}}\]
Finally, we use Stieltjes Inversion Formula Eq.~\eqref{Transforms:InvG} to obtain the asymptotic eigenvalue distribution of ${{\mathbf Q}_n}$ :
\begin{eqnarray}
{\rho}(x) = \frac{{\sqrt {\left[{{(\sqrt k  + \sqrt c )}^2} - x\right]\left[x - {{(\sqrt k  - \sqrt c )}^2}\right]} }}{{2\pi cx}}{{1}_M}(x)
\label{aa}
\end{eqnarray}
where $c=n/p$, and $M = \left({(\sqrt k  - \sqrt c )^2},{(\sqrt k  + \sqrt c )^2}\right).$

\begin{figure}[htbp]
\centering
\subfloat[ $k=5,p=200, n=200$]{\label{fig:fd1}
\includegraphics[width=0.23\textwidth]{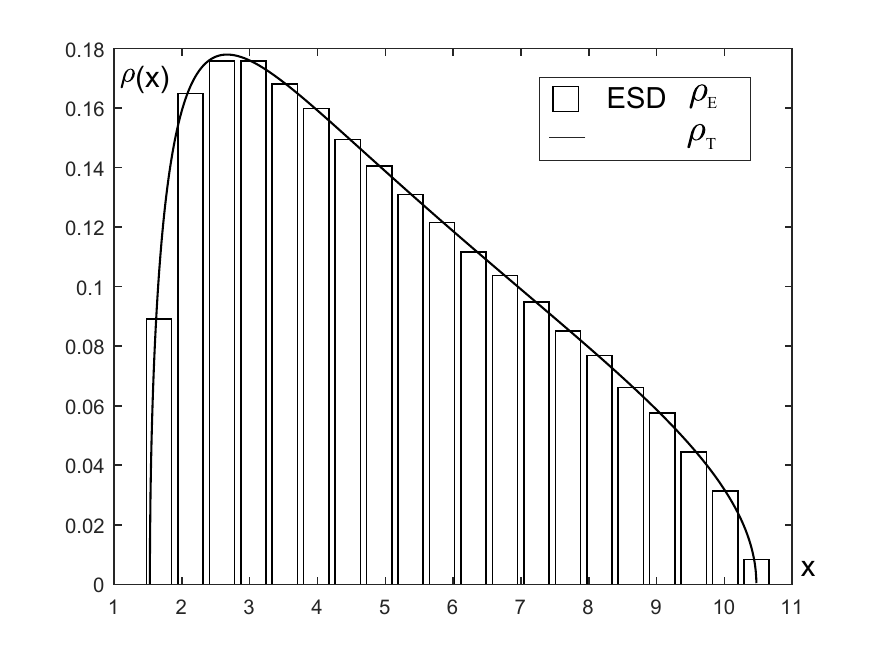}}
\subfloat[ $k=100,p=200, n=200$]{\label{fig:fd2}
\includegraphics[width=0.23\textwidth]{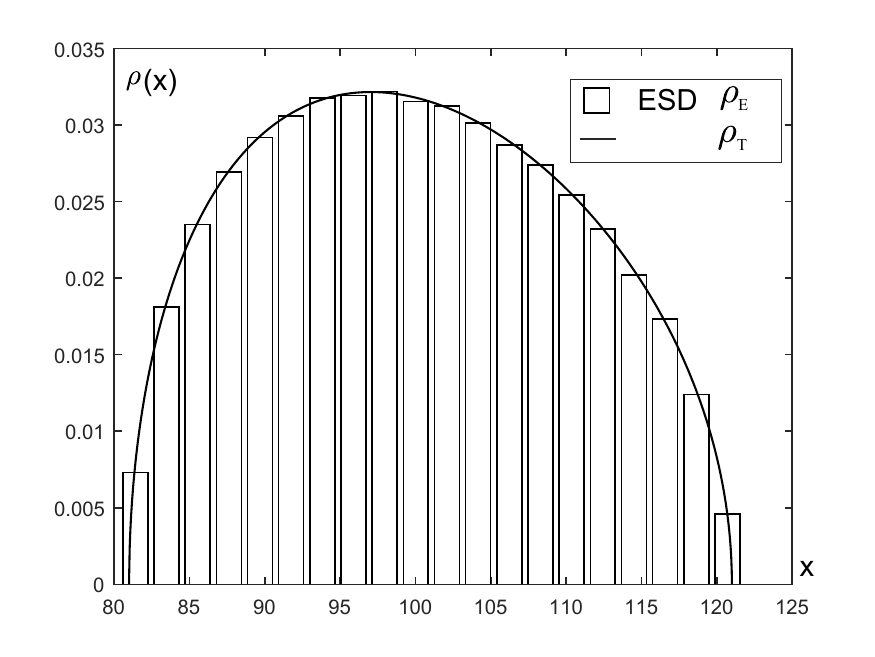}}
\caption{ESD $\bm \rho _{\text{E}}$ and theoretical $\bm \rho_{\text{T}}$ in spectrum space (Eq.~\ref{aa})}
\label{fig:fd}
\end{figure}

Figure~\ref{fig:fd} shows  the comparison of the ESD $\bm \rho _{\text{E}}$ (the bars) and theoretical asymptotic spectral density function $\bm \rho_{\text{T}}$ (the line) given in \eqref{aa}. The two are perfectly matched.

\section{Case Study in ADN}
\label{Sec: Case}

\subsection{Scene Designing and Preliminary Analysis}
\begin{figure}[htbp]
\centerline{
\includegraphics[width=.46\textwidth]{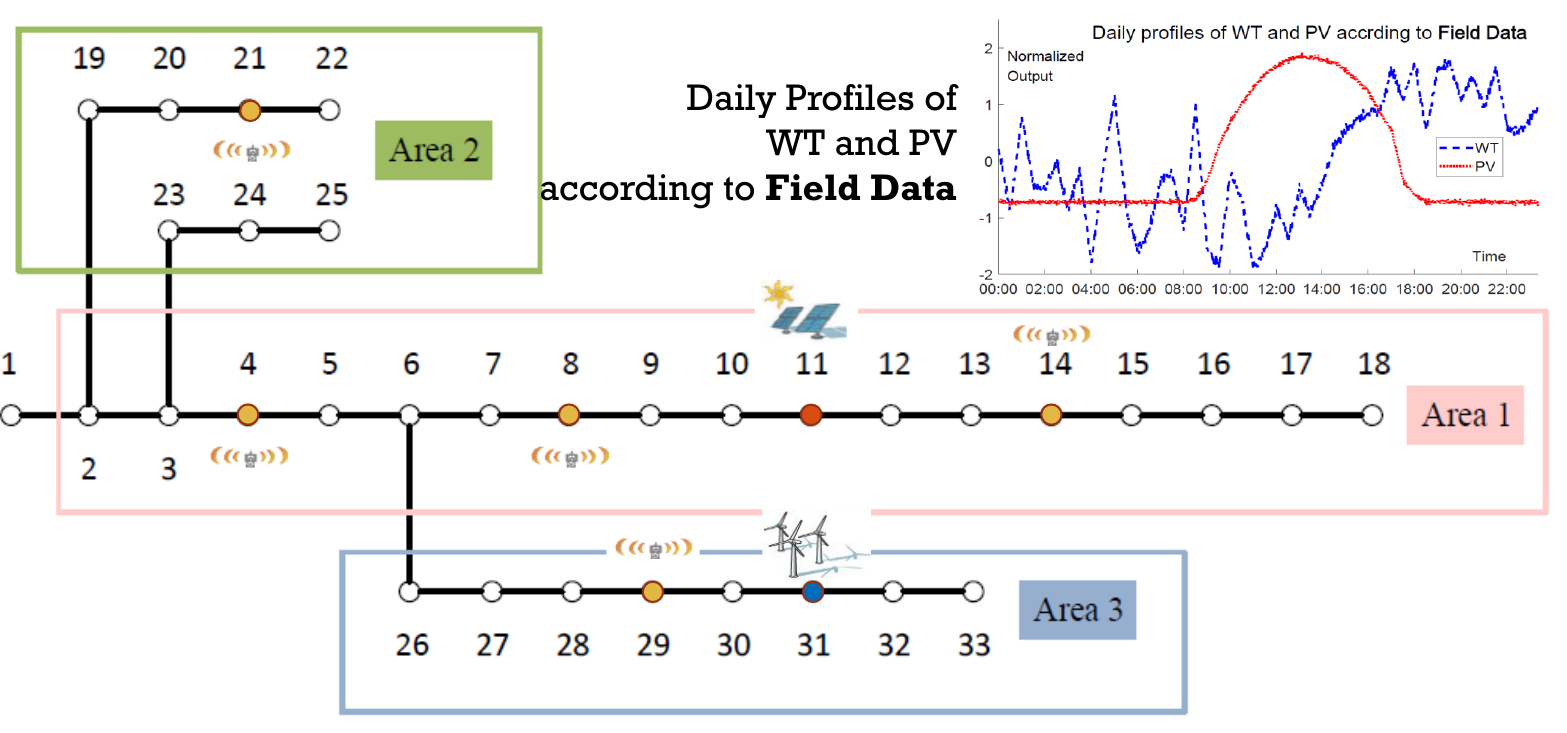}
}
\caption{Case Design: An IEEE 33-bus system with renewables}
\label{fig:Case33}
\end{figure}

Figure~\ref{fig:Case33} depicts our designed case.
We set an IEEE 33-bus distribution system with integrations of two {independent} renewables---photovoltaic and wind turbine,  respectively, on Node 11 (Area 1) and 31 (Area 3).

\subsubsection{Scene Designing}
\label{Sec: CaseDesign}
\text{\\}

Without loss of generality, two assumptions are made first:
\begin{enumerate}[a)]
\item In the mixing composite event, each event are relatively distinct. For instance, the pattern of power generated by rooftop PV, for example, is obviously different from that by generated a wind turbine.
\item Uncertainties are non-ignorable and should be factored into our task. Uncertainties are ubiquitous and play an vital role in ADN as a result of continuous injection of DER~\cite{yang2018robust}, invisible user devices~\cite{he2020invisible}, flexible network configuration~\cite{he2020hybrid}, and etc.
\end{enumerate}

The daily profiles of PV and WT are assigned to the {filed data} as Signal $\bm s_\text{PV}$ and $\bm s_\text{WT}$
according to~\cite{he2020invisible}.
We take account of noises which are independent with signals $\bm s$ on each sensor, and assume that the sampling rate is 1/min (the original sampling rate is $\frac{1}{15}$/min in practice).
We thus have a daily dataset with 1440 sampling points for further analysis.

Three scenarios are set as follows:
\begin{enumerate}[-]
\item Scene  $A$: Only PV (Signal $\bm s_\text{PV}$):$f(\mathbf X_A) = f(\mathbf S_A)\oplus f(\mathbf{R}_A)$
\item Scene  $B$: Only WT (Signal $\bm s_\text{WT}$): $f(\mathbf X_B) = f(\mathbf S_B)\oplus f(\mathbf{R}_B)$
\item Scene  $C$: PV \& WT  ($\bm s_\text{PV}$ \& $\bm s_\text{WT}$): $f(\mathbf X_C) = f(\mathbf S_C)\oplus f(\mathbf{R}_C)$
\end{enumerate}
where $\mathbf X_i, \mathbf S_i, \mathbf{R}_i\!\in\! \mathbb{R}^{33\times 1440}, i\!=\!A, B, C.$
More concretely, $\mathbf X_i$ is our observation spatial-temporal data. $\mathbf S_i$ is the signal-derived (principle) component.
$\mathbf R_i$ is the independent noise-derived component.

\begin{figure}[htbp]
\centering
\subfloat[Voltage Magnitude (Observation)]{
\includegraphics[width=0.46\textwidth]{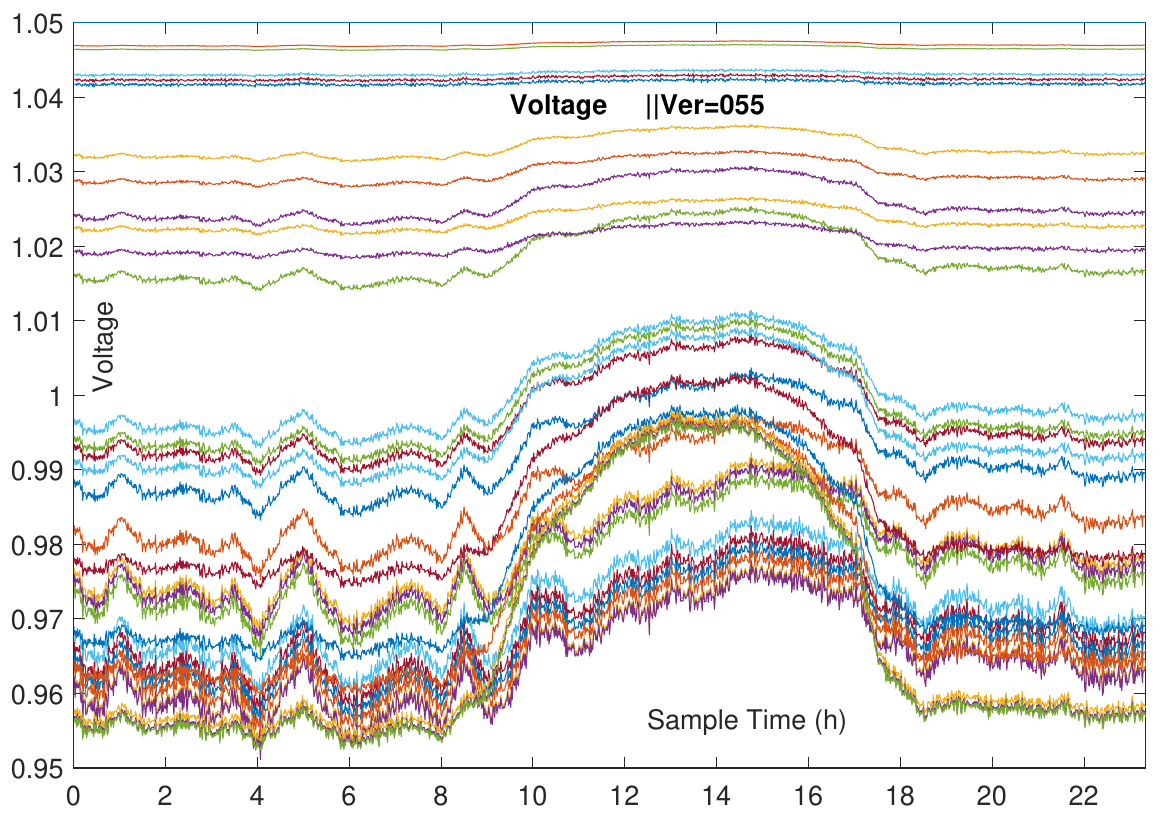}}

\subfloat[Active Power Demand (Input)]{
\includegraphics[width=0.46\textwidth]{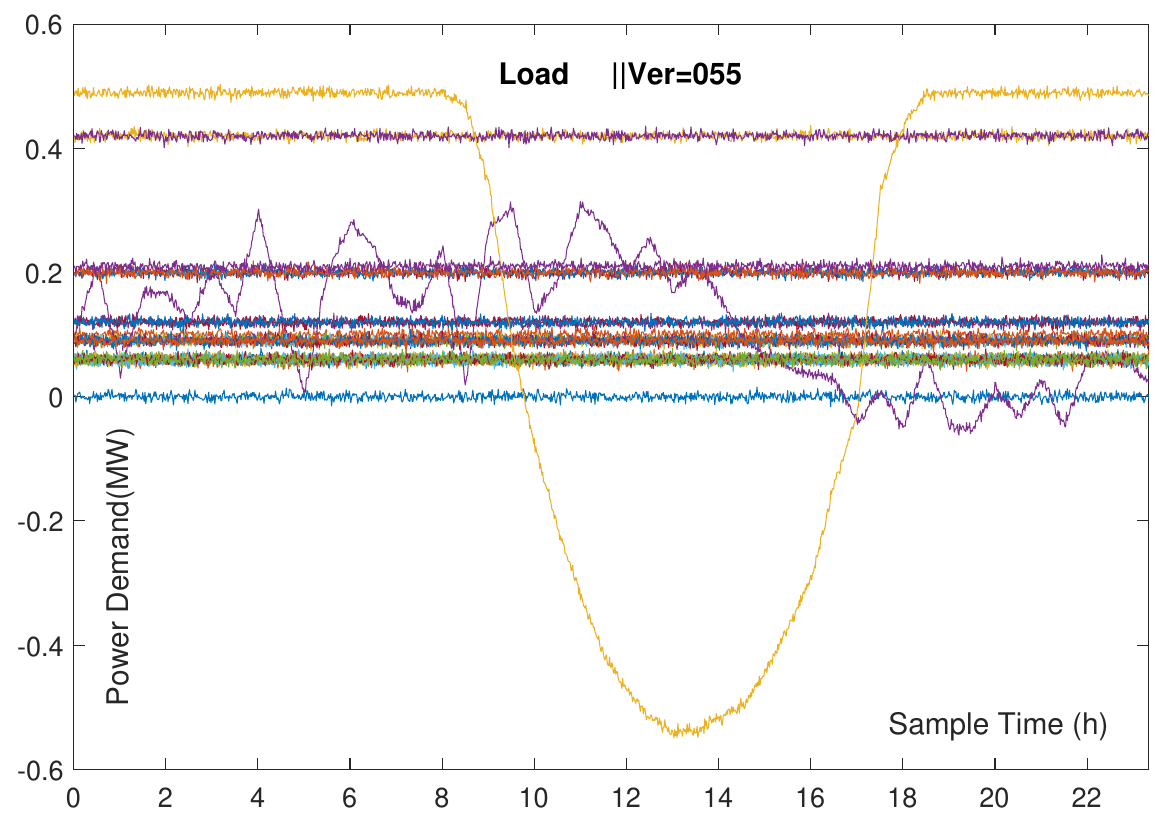}}
\caption{Outcome voltage magnitude (i.e., $\mathbf X_C$) and input power demand for Scene $C$ (WT$\oplus$PV).}
\label{fig:ObsVol}
\end{figure}

Figure~\ref{fig:ObsVol} shows the voltage magnitude (outcome) and power demand (input) for Scene $C$.
Voltage magnitude data are easily accessible to advanced sensors, and hence are chosen as spatial-temporal data $\mathbf X_C$.

\subsubsection{Spectrum Space is Immune to Non-Gaussian Noises}
\text{\\}

\begin{figure}[htbp]
\centering
\subfloat{
\includegraphics[width=0.25\textwidth]{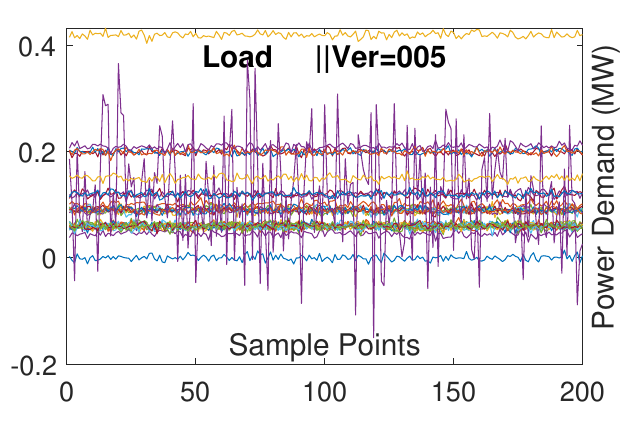}}
\subfloat{
\includegraphics[width=0.23\textwidth]{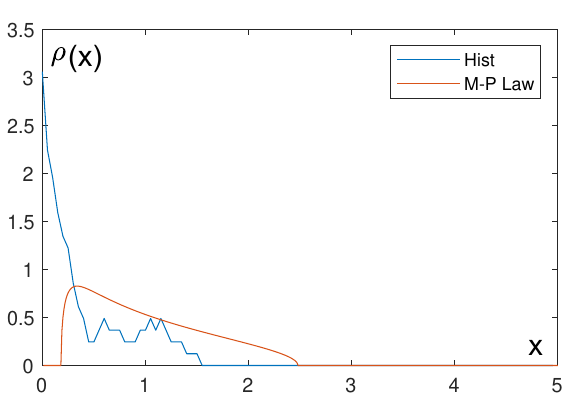}}

\subfloat{
\includegraphics[width=0.25\textwidth]{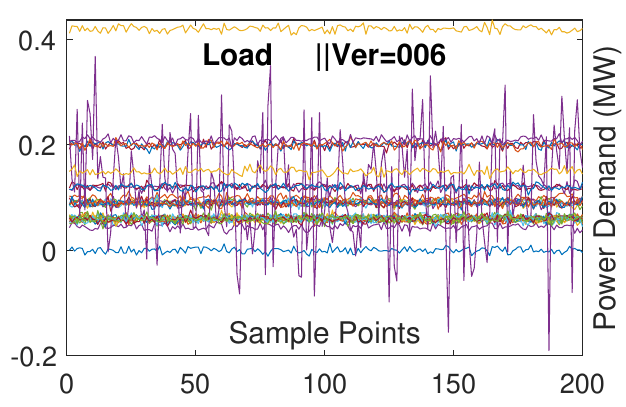}}
\subfloat{
\includegraphics[width=0.23\textwidth]{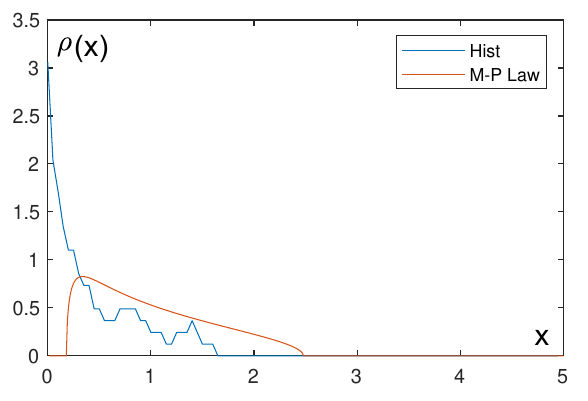}}

\subfloat{
\includegraphics[width=0.25\textwidth]{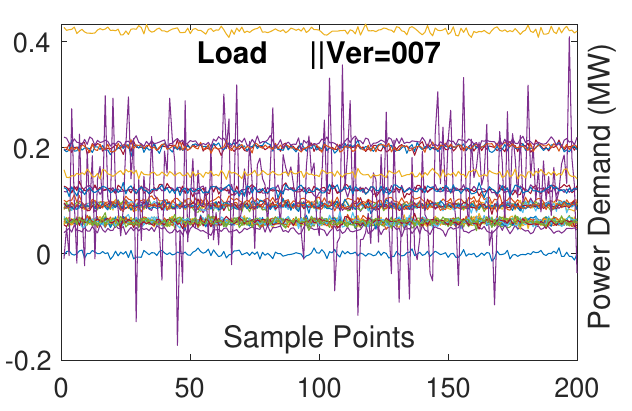}}
\subfloat{
\includegraphics[width=0.23\textwidth]{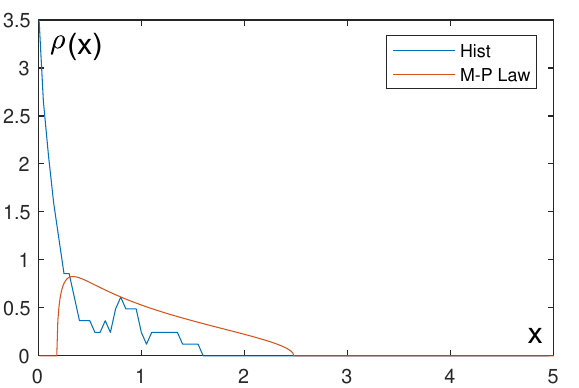}}
\caption{ESD $\bm \rho _{\text{E}}$ and theoretical density $\bm \rho_{\text{T}}$ (Eq.~\ref{aa}). Some very similar ESD $\bm \rho _{\text{E}}$ patterns are observed. {It means that the spectrum operation is immune to noises.}}
\label{fig:spectrum}
\end{figure}

The events in ADN are assumed as follows: 1) wind turbine generation on Node 31 (Event $\bar B$), and 2) regular small load fluctuations on all the nodes.
According to~\cite{huang1995use},  the auto-regression (AR) process is used to model the wind speed.
We run the simulation several times, and then obtain the voltage magnitude data $\mathbf{\Omega}$, which is easily accessible in engineering, as depicted in the left parts of Figure~\ref{fig:spectrum}.
The right parts tell that our designed mapping $f$ (i.e., $f:$ Event $\bar B$ $\mapsto$ spectral density) is immune to noise; some very similar ESD $\bm \rho _{\text{E}}$ patterns are observed in each simulation, even though wind turbine generation (following AR model) and ubiquitous noises (following i.i.d Gaussian model) are quite distinct.

\subsection{Validation of Superposition Theorem in Spectrum Space}

\begin{figure}[htbp]
\centerline{
\includegraphics[width=.46\textwidth]{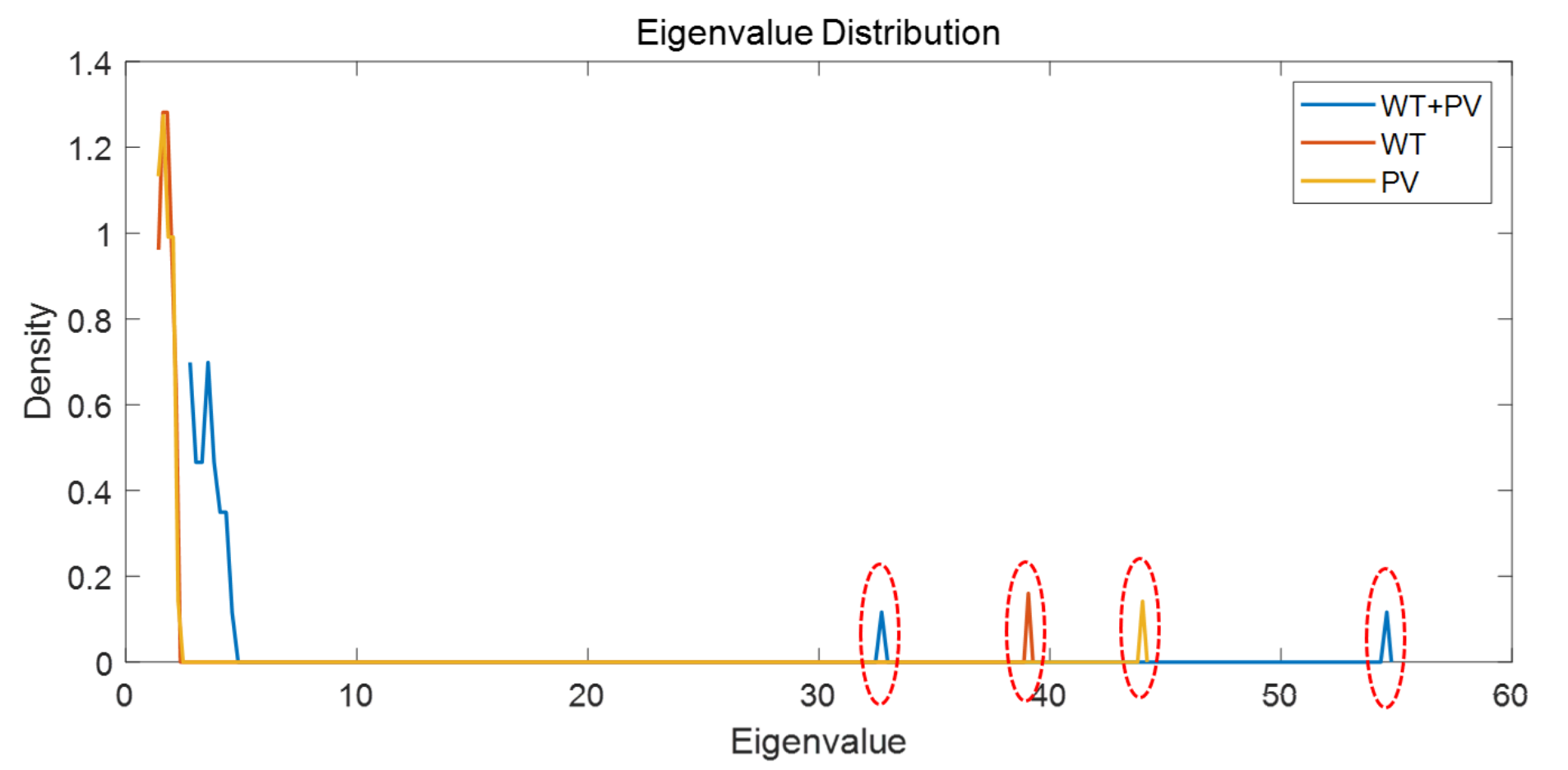}
}
\caption{Spectrum of composite event for Scene $A, B, C$}
\label{fig:CaseSpeABC}
\end{figure}

We calculate the spectrum for Scene $A, B, C$ as designed in Section~\ref{Sec: CaseDesign}.
Figure~\ref{fig:CaseSpeABC} depicts the results.
The spectrum mainly consists of two parts: A few spikes/outliers and the bulk. The former represents the part of composite event that mainly drive the features of scene, and the latter represents the part of composite event that arise from idiosyncratic noise.
It is consistent with our observation in Figure~\ref{fig:CaseSpeABC}---one outlier for Scene $A$, one for Scene $B$, and two for Scene $C$.

\begin{figure}[htbp]
\centering
\subfloat[Real Part of $G(z)$]{\label{fig:CaseGtran1}
\includegraphics[width=0.48\textwidth]{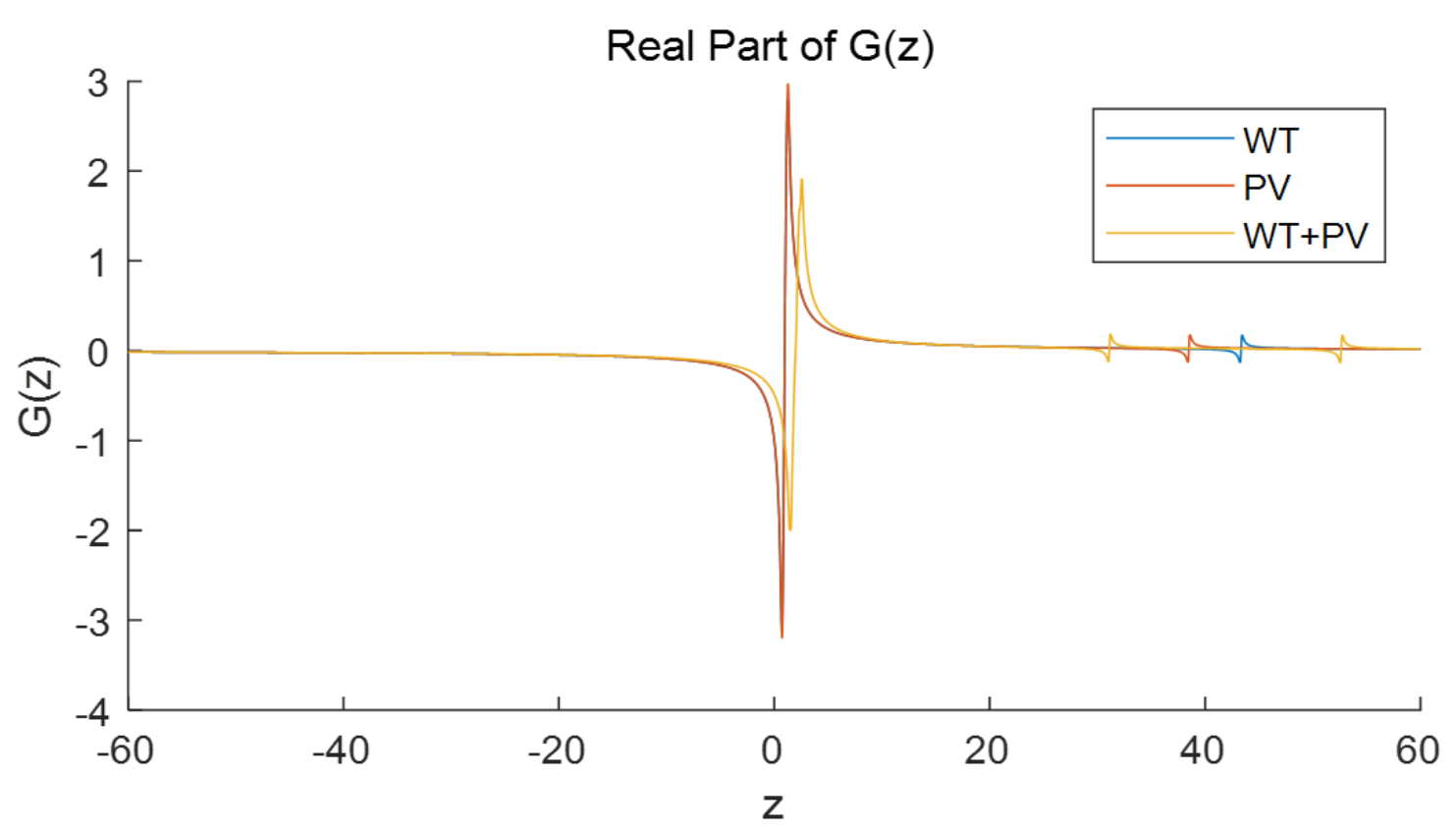}}

\subfloat[Imag Part of $G(z)$]{\label{fig:CaseGtran2}
\includegraphics[width=0.48\textwidth]{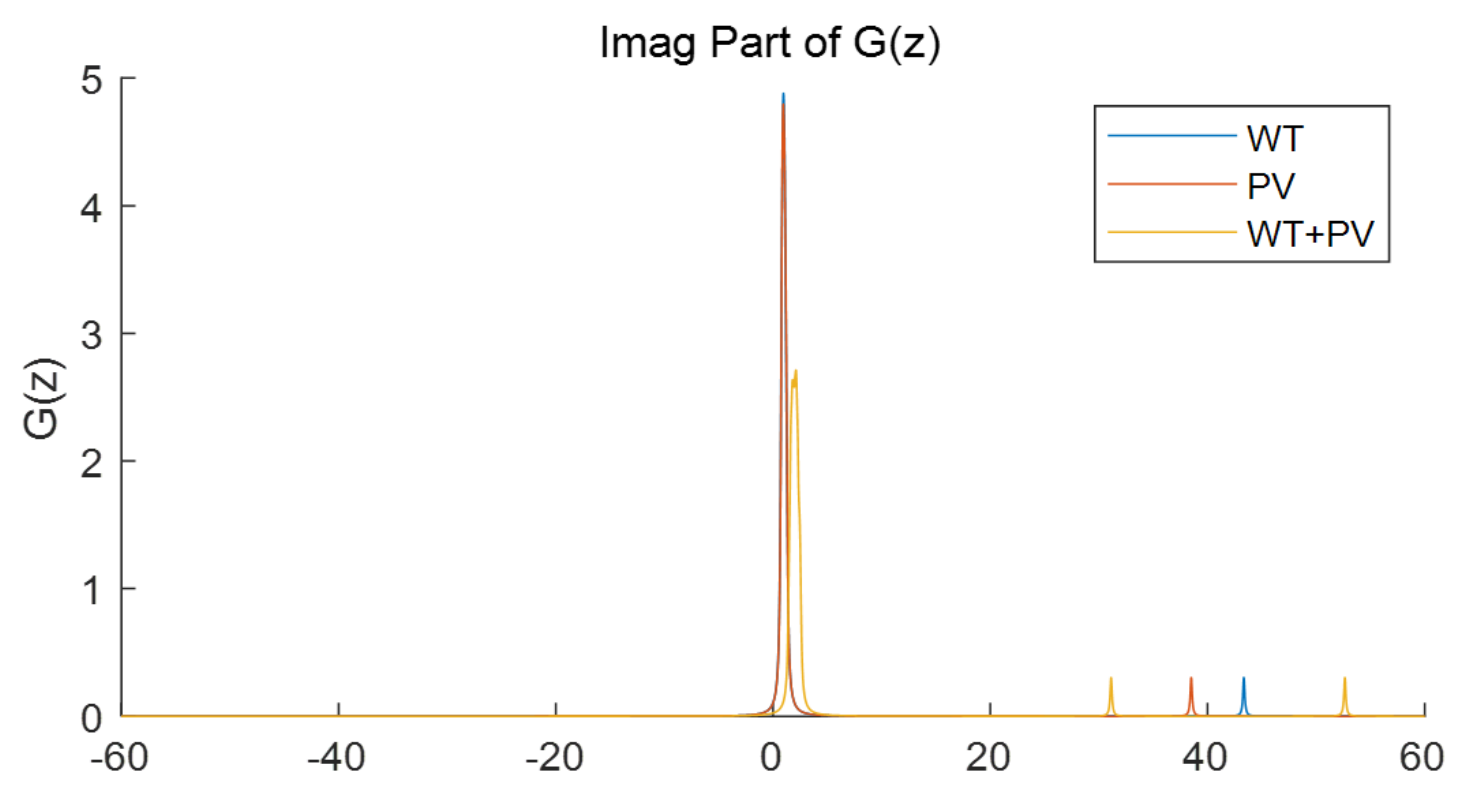}}
\caption{Numerical solution of Stieltjes transform $G(z)$}
\label{fig:CaseGtran}
\end{figure}

According to Section~\ref{Sec: STFor}, the events addition is formulated in the spectrum space.
With the field data, we ordinarily can not build the exact analytic expression for the eigenvalue density $\rho$.
In practice, instead of  analytical solution as Eq.~\eqref{eq:STJtrans},  we acquire $G(z)$ through numerical solution.
Figure~\ref{fig:CaseGtran} depicts the resultant outcomes of numerical solution.

\begin{figure}[htbp]
\centerline{
\includegraphics[width=.46\textwidth]{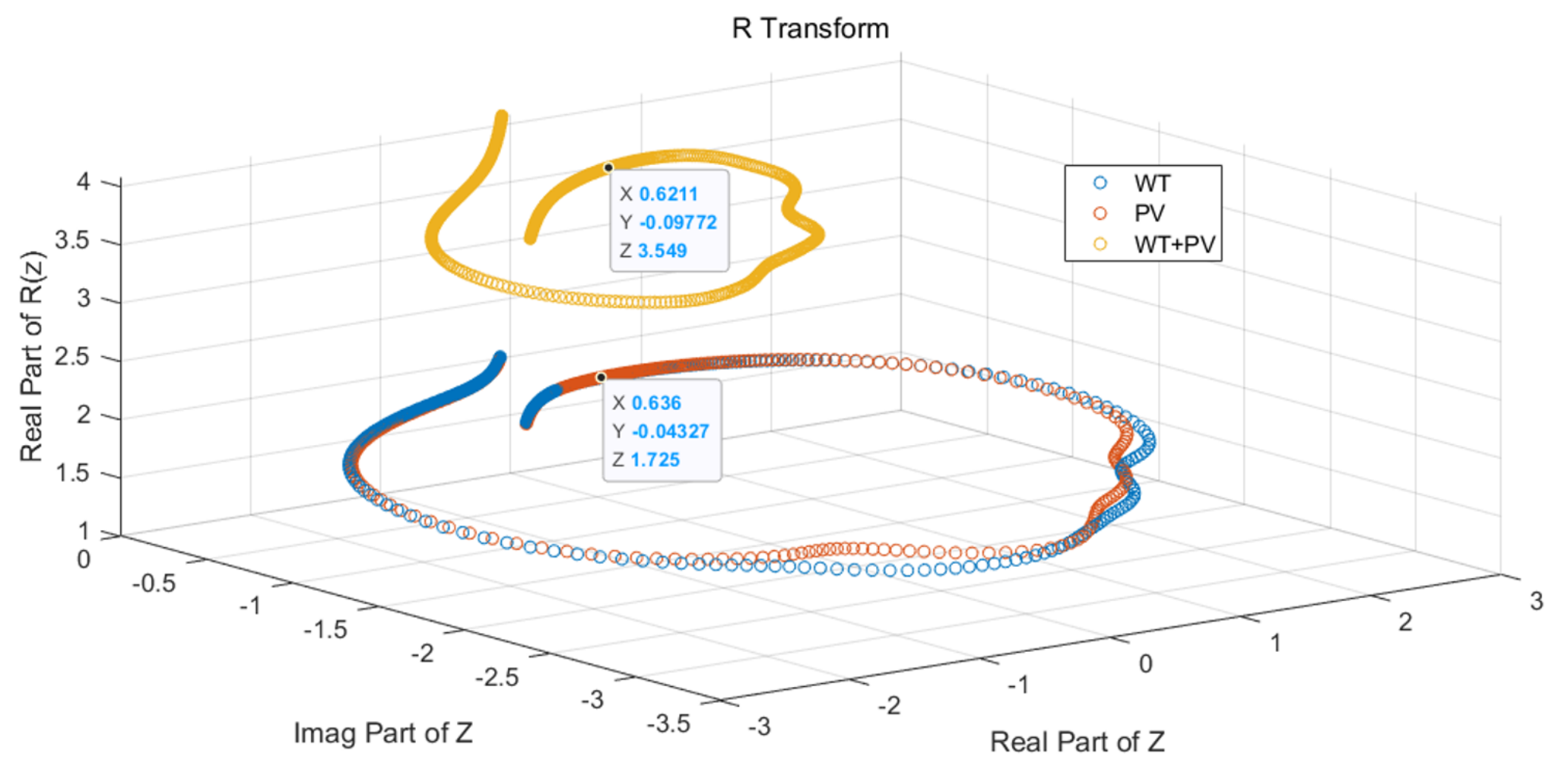}
}
\caption{Resultant outcomes of R transform $R(z)$}
\label{fig:CaseRtran}
\end{figure}

According to the real part of $G(z)$ (Figure~\ref{fig:CaseGtran1}), R transform is implemented among $z\in \left[ -3+i\varepsilon ,3+i\varepsilon  \right]$.
The result is depicted in Figure~\ref{fig:CaseRtran}.
It gives a rough display of $R_C=R_A\oplus R_B$. For instance, when $z\approx0.62-i0.098$, it is found that  $R_A = 1.725$,  $R_B = 1.728$, and $R_C = 3.549 \approx R_A + R_B$  (where $R_i$ means  real$(R_i(z)), i = A, B, C)$.

\begin{figure*}[htbp]
\centering
\subfloat[$\sum{\left|R_C-2R_A\right|=4897.2}$]{\label{fig:CaseRes1}
\includegraphics[width=0.32\textwidth]{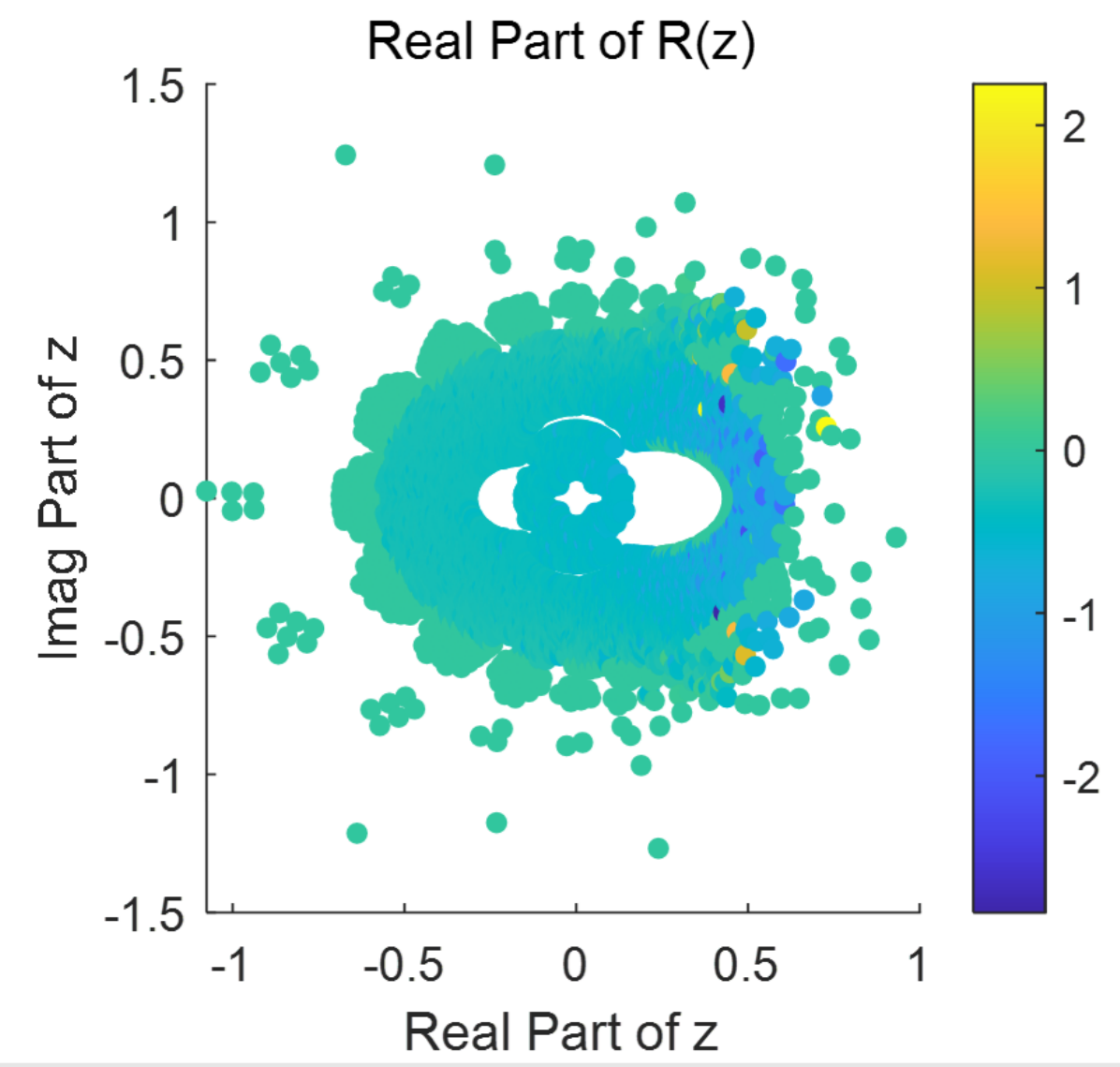}}
\subfloat[$\sum{\left|R_C-2R_B\right|=5120.3}$]{\label{fig:CaseRes2}
\includegraphics[width=0.32\textwidth]{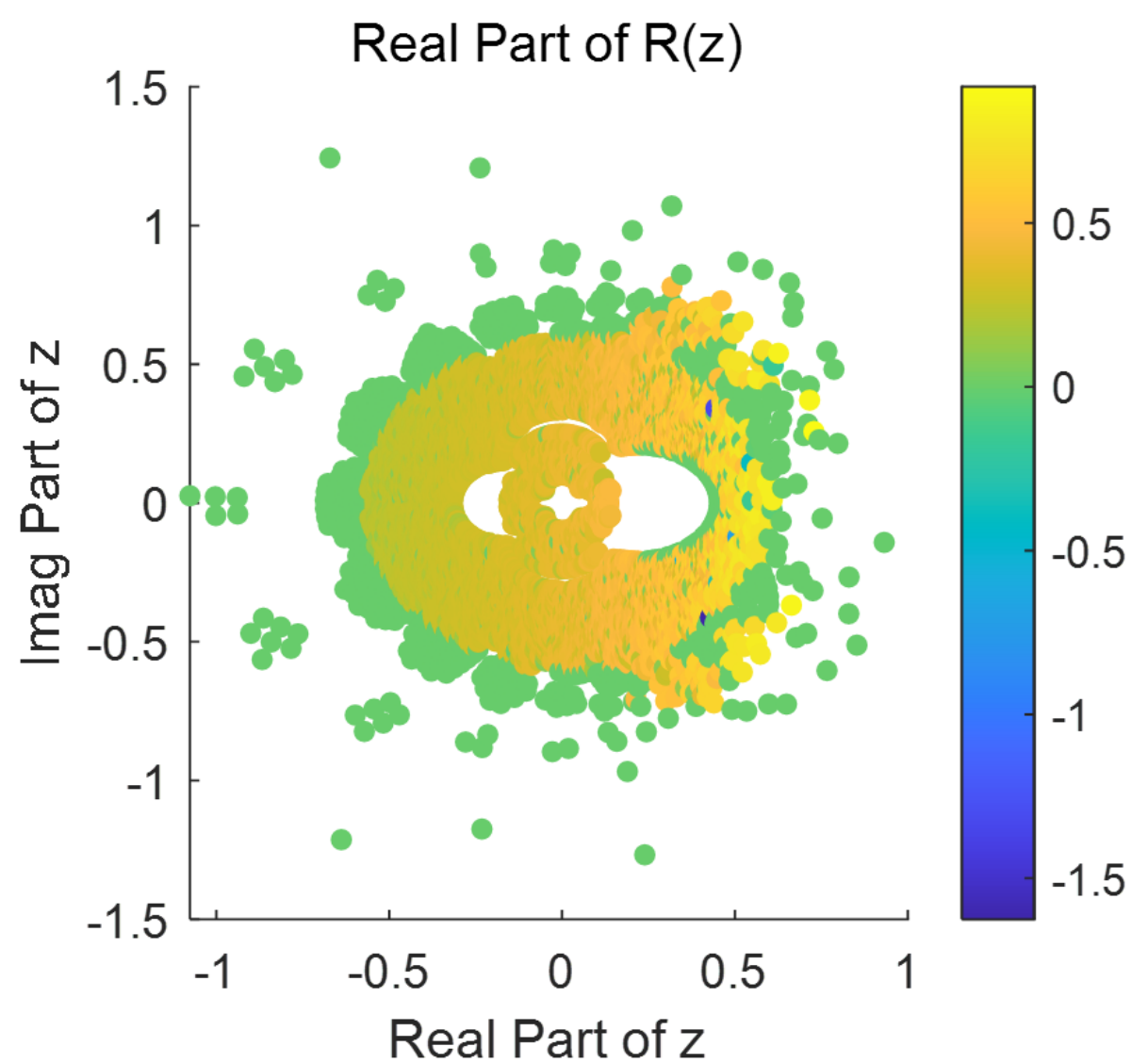}}
\subfloat[$\sum{\left|R_C-R_A-R_B\right|}=405.1$]{\label{fig:CaseRes3}
\includegraphics[width=0.32\textwidth]{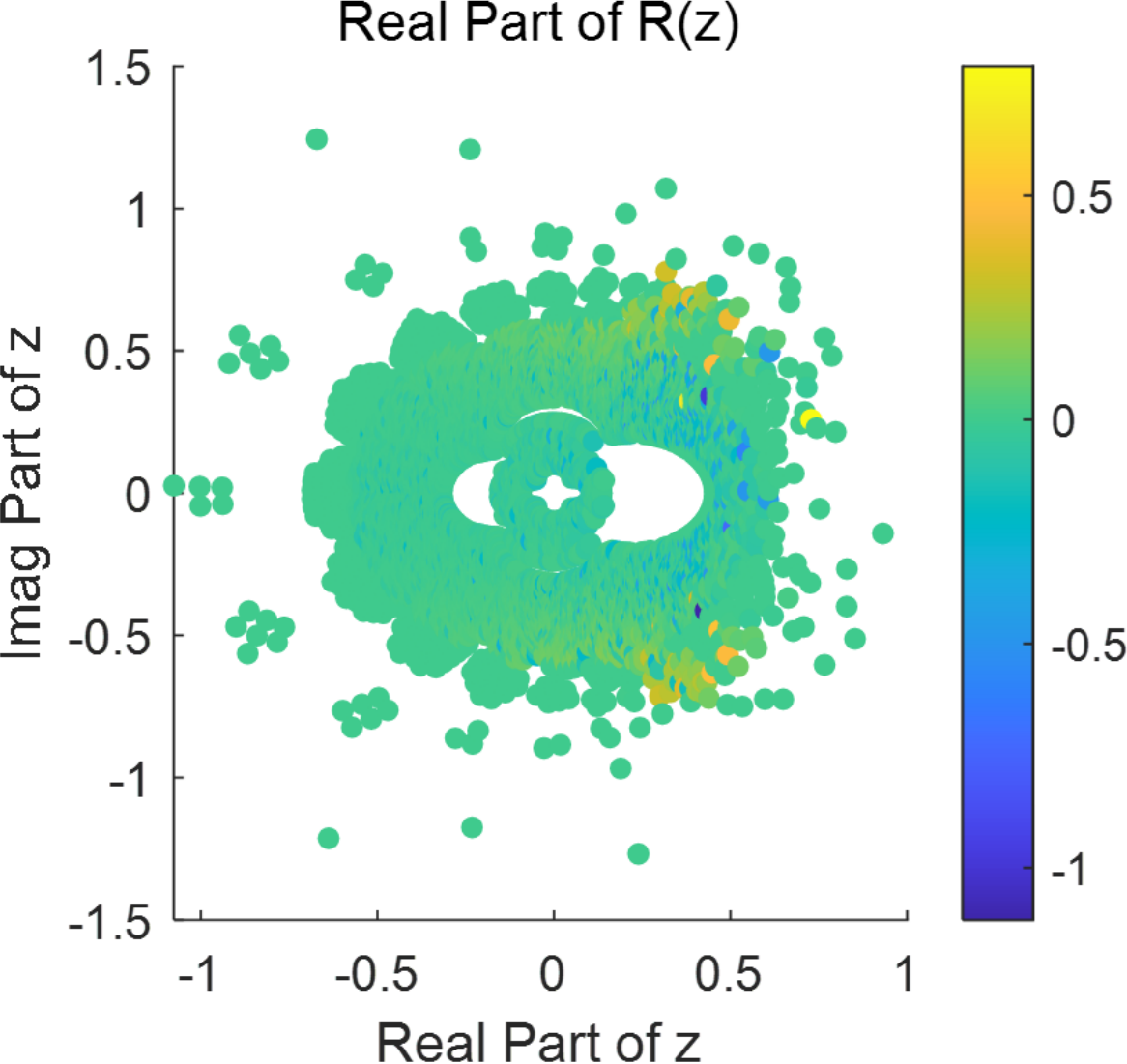}}
\caption{Sum of absolute residue. $R_C=R_A\oplus R_B$ in spectrum space does make sense.}
\label{fig:CaseRABC}
\end{figure*}

Furthermore, we implement test on $R_C-R_i-R_j$ (i.e., $R_C-2R_A$, $R_C-2R_B$, and $R_C-R_A-R_B$), and Figure~\ref{fig:CaseRABC} depicts the results respectively.
We then sum their absolute value, i.e., $\sum{\left|R_C-R_i-R_j\right|}$, and obtain $4897.2, 5120.3,$ and $405.1$ for each case.
$R_C-R_A-R_B$ is significantly superior to the other two combinations (\textbf{at $10^1$ level}), which means the composite event is more likely to be composed of A and B (i.e., $f(C) = f(A)\oplus f(B)$).
{In practice, $R_C$, based on our observation, is calculable, while $R_A$ and $R_B$ are experience-based and immune to noises in the spectrum space (see Figure~\ref{fig:spectrum}).
This case verifies that the RMT-derived addition $\oplus$ does help composite event analysis in ADN.
}

\section{Conclusion}
\label{Sec: Conc}
\normalsize{}
In conclusion, this study has presented a groundbreaking formulation of the Superposition Theorem (ST) within the spectrum domain. Tying together Stieltjes transform and R transform, we unveil spectrum-based insights into the observed composite event, allowing us to conduct analysis on those previously invisible atom events in an active distribution network (ADN). These insights provide a novel framework for modeling the behaviors of Distributed Energy Resources (DER) integrated across the ADN.

A pivotal achievement of this study is the development of the formulation denoted as  $f(\mathbf M)=f(\mathbf A_1)\oplus f(\mathbf A_2)\oplus\cdots$, which characterizes the superposition behavior of composite events.
This formulation is model-free, non-linear, non-supervised, theory-guided, and uncertainty-insensitive.
Its validation using field data from the IEEE-33 bus system further demonstrates its applicability.
By providing an abstract ``$\oplus$'' based ST in spectrum space, this study offers valuable insights for network dispatchers and operators.
It holds the potential to advance the analysis of composite events within the ADN, potentially yielding rules akin to Th\'evenin Conversion for nonlinear analysis.
Besides, this study may benefit its inversion, blind decomposition of composite event (BDCE), and thus some advanced tasks, such as the extraction and identification of characteristic components related to faults or anomalies.

The aspiration to merge traditional tasks in power systems with cutting-edge tools in data science represents a long-term goal within our community. This study represents a significant step towards realizing this vision, highlighting the possibilities for innovative advancements at the intersection of power systems and data science. It underscores the potential for harnessing data-driven approaches to enhance the reliability, resilience, and efficiency of modern power grids.

\begin{appendices}

\section{Stieltjes Transform of GUE (Eq.~\ref{eq:STJtrans})}
\label{Sec:ProofofStj}

Recall Eq.~\eqref{eq:D2} $\mathrm{d} v(t)=(2 \pi)^{-1} \sqrt{4-t^{2}}$, $t\in \left[ -2,2 \right]$.
Thus
$$G(z)=\frac{1}{2 \pi} \int_{-2}^{2} \frac{\sqrt{4-t^{2}}}{z-t} \mathrm{d} t$$

(\expandafter{\romannumeral1}) Make the substitution $t=2 \cos \theta$ for $0 \leq \theta \leq \pi$:
$$G(z)=\frac{1}{2\pi }\int_{\pi }^{0}{\frac{2\sin \theta }{z-2\cos \theta }\text{d}2\cos \theta }=\frac{1}{4\pi }\int_{0}^{2\pi }{\frac{4{{\sin }^{2}}\theta }{z-2\cos \theta }}\text{d}\theta $$

(\expandafter{\romannumeral2}) Make the substitution $t=2 \cos \theta$ for $0 \leq \theta \leq \pi$:
\small
\[\begin{aligned}
   G(z)&=\frac{1}{4\pi }\int_{0}^{2\pi }{\frac{4{{\sin }^{2}}\theta }{z-2\cos \theta }\text{d}\theta }=\frac{1}{4\pi }\oint_{C}{\frac{{{\left( -i\left( w-1/w \right) \right)}^{2}}}{z-w+1/w}\frac{1}{wi}\text{d}w} \\
 & =\frac{1}{4\pi i}\oint_{C}{\frac{{{\left( {{w}^{2}}-1 \right)}^{2}}}{{{w}^{2}}\left( {{w}^{2}}-zw+1 \right)}\text{d}w}
\end{aligned}\]
\normalsize
where $\Gamma=\{w \in \mathbb{C}| \left|{w}\right| = 1\} .$

Tips:
\[\begin{aligned}
  & w={{e}^{i\theta }}=\cos \theta +i\sin \theta \\
  &\text{d}w=wi\text{d}\theta  \\
  &1/w=1/{{e}^{i\theta }}={{e}^{-i\theta }}=\cos \theta -i\sin \theta  \\
 & \left\{ \begin{aligned}
  & 2\cos \theta =w+1/w \\
 & 2\sin \theta =-i\left( w-1/w \right) \\
\end{aligned} \right. \\
\end{aligned}\]

(\expandafter{\romannumeral3}) Show that the roots of $ {{w}^{2}}-zw+1=0$ are $w_{1}=\left(z-\sqrt{z^{2}-4}\right) / 2$ and $w_{2}=\left(z+\sqrt{z^{2}-4}\right) / 2$. $w_{1} \in \operatorname{int}(\Gamma)$ and $w_{2} \notin \operatorname{int}(\Gamma)$, using the branch defined above.

Tips:

For $z \in \mathbb{C}^{+}$ (i.e., $z\in \mathbb C$, Im$(z)>0$)
\[\begin{aligned}
&\left|\operatorname{Im}(z)\right|<\left|\operatorname{Im}\left(\sqrt{z^{2}-4}\right)\right| \\
&\left|\operatorname{Re}\left(\sqrt{z^{2}-4}\right)\right| \leq\left|\operatorname{Re}(z)\right| \\
\end{aligned}\]
with equality in the second relation only when $\operatorname{Re}(z)=0$.

(\expandafter{\romannumeral4})  Using the residue calculus, show that:
\[G(z)=\frac{z-\sqrt{z^{2}-4}}{2}\].

\section{Proof of Stieltjes Inversion Formula (Eq.~\ref{Transforms:InvG})}
\label{Sec: InvStj}
\textit{Proof:}  We have
\[
\small{}
\begin{aligned}
\operatorname{Im}(G(x+i y))&=\operatorname{Im}\int_{\mathbb{R}} \left(\frac{1}{x-s+i y}\right) \text d v(s)\\
                                          &=\int_{\mathbb{R}} \operatorname{Im}\left(\frac{1}{x-s+i y}\right) \text d v(s)\\
                                          &=\int_{\mathbb{R}} \frac{-y}{(x-s)^{2}+y^{2}} d v(s)
\end{aligned}
\normalsize{}
\]
Thus
$$
\small{}
\begin{aligned}
 \int_{a}^{b} \operatorname{Im}(G(x+i y)) d x &=\int_{a}^{b}\int_{\mathbb{R}} \frac{-y}{(x-s)^{2}+y^{2}} d v(s) d x  \\
&= \int_{\mathbb{R}} \int_{a}^{b}\frac{-y}{(x-s)^{2}+y^{2}} d x d v(s) \\
&=-\int_{\mathbb{R}} \int_{(a-1) / y}^{(b-s) / y} \frac{1}{1+\tilde{x}^{2}} d \tilde{x} d v(s) \\
&=-\int_{\mathbb{R}}\left[\arctan  \left. {\tilde x}  \right|_{(a-1) / y}^{(b-1) / y}\right] d v(s)
\end{aligned}
\normalsize{}
$$

Let $f(y, t)=\arctan((b-s) / y)-\arctan ((a-s) / y)$, and
$$
\small{}
\lim_{y \rightarrow 0^{+}}f(y, s)=\left\{\begin{array}{ll}
0, & s \notin[a, b] \\
\pi / 2, & s \in\{a, b\} \\
\pi, & s \in(a, b)
\end{array}\right.
\normalsize{}
$$

As a result
$$
\small
-\!\lim_{y \rightarrow 0^{+}} \frac{1}{\pi} \int_{a}^{b} \operatorname{Im}(G(x+i y)) d x\!=\!v((a, b))\!+\!\frac{1}{2} v(\{a\}) \!+\!\frac{1}{2} v(\{b\})
\normalsize
$$
put $a = x_0, b = x_0\!+\!\Delta x$, $\epsilon \to 0^{+}$
$$
\small
-\frac{1}{\pi} \int_{x_0}^{x_0\!+\!\Delta x} \operatorname{Im}(G(x+i\epsilon)) d x\!=\!v((x_0,x_0+\Delta x))
\normalsize
$$
$$
\Rightarrow -\frac{1}{\pi}\operatorname{Im}(G(x_0+i\epsilon))\Delta x  = \rho(x_0)\Delta x
$$
$$
\Rightarrow \rho(x_0) = -\frac{1}{\pi}\operatorname{Im}(G(x_0+i\epsilon)),  \  \epsilon\to 0^{+}
$$

\section{Relationship of Moment and Cumulant}
\label{Sec:Bell}
The definition of Moment and Cumulant tells that
\[
M\left( t \right):=\mathbb{E}\left( {{e}^{Xt}} \right)=\mathbb{E}\left( \sum\limits_{k=0}^{\infty }{{{X}^{k}}\frac{{{t}^{k}}}{k!}} \right) =  \sum\limits_{k=0}^{\infty }{\frac{{{t}^{k}}}{k!}\mathbb{E}\left({X}^{k}\right)}
\]

\[ M\left( t \right)={{e}^{K\left( t \right)}}
  \Rightarrow \sum\limits_{k=0}^{\infty }{{{m}_{k}}\frac{{{t}^{k}}}{k!}}=\sum\limits_{n=0}^{\infty }{\frac{{{K}^{n}}\left( t \right)}{n!}}
\]

Check the $n$-th term $t^n$ of both side

\[\begin{aligned}
  \text{L}&=\sum\limits_{k=0}^{\infty }{{{m}_{k}}\frac{{{t}^{k}}}{k!}}\rightarrow{{m}_{n}}\frac{1}{n!}{{t}^{n}} \\
  \text{R}&=\frac{{{K}^{1}}\left( t \right)}{1!}\!+\!\frac{{{K}^{2}}\left( t \right)}{2!}+\cdots +\frac{{{K}^{n}}\left( t \right)}{n!}+\cdots=\sum\limits_{k=1}^{n}{\frac{{{K}^{k}}\left( t \right)}{k!}}  \\
  & =\sum\limits_{k=1}^{n}{\frac{{{\left( {{\kappa }_{1}}t+\frac{{{\kappa }_{2}}}{2!}{{t}^{2}}+\frac{{{\kappa }_{3}}}{3!}{{t}^{3}}+\cdots +\frac{{{\kappa }_{n}}}{n!}{{t}^{n}}+\cdots  \right)}^{k}}}{k!}}\\
  &\rightarrow \sum{\frac{1}{{k!}}\frac{{k!}}{{{j}_{1}}!{{j}_{2}}!\cdots {{j}_{n}}!}{{\left( \frac{{{\kappa }_{1}}}{1!} \right)}^{{{j}_{1}}}}{{\left( \frac{{{\kappa }_{2}}}{2!} \right)}^{{{j}_{2}}}}\cdots {{\left( \frac{{{\kappa }_{n}}}{n!} \right)}^{{{j}_{n}}}}}{t}^{n}
\end{aligned}\]
where
\[\left\{ \begin{aligned}
  & {{j}_{1}},{{j}_{2}},\cdots +{{j}_{n}}\ge 0 \\
 & {{j}_{1}}+2{{j}_{2}}+\cdots +n{{j}_{n}}=n \\
 & {{j}_{1}}+{{j}_{2}}+\cdots +{{j}_{n}}=k=\#\pi \le n \\
\end{aligned} \right.\]

As a result
\[\begin{aligned}
   {{m}_{n}}\frac{1}{n!}&=\sum{\frac{1}{{k!}}\frac{{k!}}{{{j}_{1}}!{{j}_{2}}!\cdots {{j}_{n}}!}{{\left( \frac{{{\kappa }_{1}}}{1!} \right)}^{{{j}_{1}}}}{{\left( \frac{{{\kappa }_{2}}}{2!} \right)}^{{{j}_{2}}}}\cdots {{\left( \frac{{{\kappa }_{n}}}{n!} \right)}^{{{j}_{n}}}}}\\
  {{m}_{n}}&=\sum{\frac{n!}{{{j}_{1}}!{{j}_{2}}!\cdots {{j}_{n}}!}{{\left( \frac{{{\kappa }_{1}}}{1!} \right)}^{{{j}_{1}}}}{{\left( \frac{{{\kappa }_{2}}}{2!} \right)}^{{{j}_{2}}}}\cdots {{\left( \frac{{{\kappa }_{n}}}{n!} \right)}^{{{j}_{n}}}}}\\
 &=\sum\limits_{\pi \in \mathcal{P}\left( \left[ n \right] \right)}{{{B}_{n,k}}\left( \kappa  \right)}={{B}_{n}}\left( \kappa  \right)
\end{aligned}\]

where
\[{{B}_{n,k}}\left( {\kappa } \right)=\sum{\frac{n!}{{{j}_{1}}!{{j}_{2}}!\cdots {{j}_{n}}!}{{\left( \frac{{{\kappa }_{1}}}{1!} \right)}^{{{j}_{1}}}}{{\left( \frac{{{\kappa }_{2}}}{2!} \right)}^{{{j}_{2}}}}\cdots {{\left( \frac{{{\kappa }_{n}}}{n!} \right)}^{{{j}_{n}}}}}\]
\[
\begin{aligned}{{B}_{n}}\left( {{\kappa }} \right)&=\sum\limits_{k=0}^{n}{{{B}_{n,k}}\left( {{\kappa }} \right)}\\
&=\sum{\frac{n!}{{{j}_{1}}!{{j}_{2}}!\cdots {{j}_{n}}!}{{\left( \frac{{{\kappa }_{1}}}{1!} \right)}^{{{j}_{1}}}}{{\left( \frac{{{\kappa }_{2}}}{2!} \right)}^{{{j}_{2}}}}\cdots {{\left( \frac{{{\kappa }_{n}}}{n!} \right)}^{{{j}_{n}}}}}\\
\end{aligned}
\]

and thus ${{m}_{n}}=\sum\limits_{\pi \in \mathcal{P}\left( \left[ n \right] \right)}{{{\kappa }_{\pi }}}={{B}_{n}}\left( {{\kappa }} \right).$

\section{Proof of $\xi(z)=1+\zeta(z \xi(z))$ (Eq.~\ref{eq:MomentSeriesCumulantSeries})}
\label{Sec:MzKz}

\textit{Proposition 1:}

\textit{Proof:}
\[\text{L}:=\xi(z)=1+\zeta(z \xi(z)):=\text{R}\]

Check the $n$-th derivative $f^{(n)}(0)$ of both side
\begin{flalign}
&\   \text{L}=\xi(z)=\sum\limits_{k=0}^{\infty }{{{m}_{k}}{z}^{k}} \rightarrow n!m_n  \nonumber \\
&\
\begin{aligned}
\text{R}&=1+\zeta(z \xi(z))= 1+\kappa_1(z \xi(z))+\kappa_2(z^2 \xi^2(z))+\cdots\\
\end{aligned}
\nonumber &
\end{flalign}

\section{Proof of $G^{-1}(z)$ (Eq.~\ref{eq:T_invG})}
\label{Sec:ProofofG-1}
Put $T(z)\!=\!R(z)+\frac{1}{z}\overset{\text{Eq}\text{.}\,\left( \text{15} \right)}{\mathop{=}}\,\frac{K(z)+1}{z} .$ Then we have
$$
\small
\begin{aligned}
   T\left( G\left( z \right) \right)&=\frac{K\left( G\left( z \right) \right)+1}{G\left( z \right)}\overset{\text{Eq}\text{.}\,\left( \text{14} \right)}{\mathop{=}}\,\frac{1}{G\left( z \right)}\left(K\left( \frac{1}{z}M\left( \frac{1}{z} \right) \right)+1\right) \\
 & \overset{\text{Eq}\text{.}\,\left( \text{13} \right)}{\mathop{=}}\frac{1}{G(z)}M\left( \frac{1}{z} \right)\overset{\text{Eq}\text{.}\,\left( \text{14} \right)}{\mathop{=}}\,\frac{1}{G(z)}zG(z)=z
\end{aligned}
\normalsize
$$
Thus $G^{-1}(z)=T(z)=R(z)+\frac{1}{z}$

\end{appendices}

\bibliographystyle{IEEEtran}
\bibliography{helx}

\normalsize{}
\end{document}